RESEARCH ARTICLE

# *Inforence*: Effective Fault Localization Based on Information-Theoretic Analysis and Statistical Causal Inference

Farid FEYZI[1], Saeed PARSA[1] (✉)

1 Department of Computer Engineering, Iran University of Science and Technology, Tehran, Iran

**Abstract** In this paper, a novel approach, *Inforence,* is proposed to isolate the suspicious codes that likely contain faults. *Inforence* employs a feature selection method, based on mutual information, to identify those bug-related statements that may cause the program to fail. Because the majority of a program faults may be revealed as undesired joint effect of the program statements on each other and on program termination state, unlike the state-of-the-art methods*, Inforence* tries to identify and select groups of interdependent statements which altogether may affect the program failure. The interdependence amongst the statements is measured according to their mutual effect on each other and on the program termination state. To provide the context of failure, the selected bug-related statements are chained to each other, considering the program static structure. Eventually, the resultant cause-effect chains are ranked according to their combined causal effect on program failure. To validate *Inforence,* the results of our experiments with seven sets of programs include Siemens suite, gzip, grep, sed, space, make and bash are presented. The experimental results are then compared with those provided by different fault localization techniques for the both single-fault and multi-fault programs. The experimental results prove the outperformance of the proposed method compared to the state-of-the-art techniques.

**Keywords** Fault Localization, Debugging, Backward Dynamic Slice, Mutual Information, Feature Selection

## 1 Introduction

To eliminate a bug, programmers employ all means to identify the location of the bug and figure out its cause. This process is referred to as software fault localization, which is one of the most expensive activities of debugging. Due to intricacy and inaccuracy of manual fault localization, a great amount of research has been carried out to develop automated techniques and tools to assist developers in finding bugs [1-11]. Most of these techniques use dynamic information from test executions, known as Spectrum-based Fault Localization (SBFL).

The majority of SBFL techniques do not perform well in the case of specific bugs caused by undesired interactions between statements because they only consider statements in isolation. In other words, for each individual statement, they contrast its presence in all failing and passing runs to assign a fault suspiciousness value according to the contrast measure. However, as shown in Section 2, there are certain situations in which a specific combination of statements causes undesired program results. Hence, modeling the combinatorial effect of statements on each other, in failing and passing executions, may considerably improve the fault localization process. In this regard, the new idea of locating failure-causing statements considering their combinatorial effect on the program failure is suggested. The idea is inspired by the observation that most program failures are only revealed when a specific combination of correlated statements are executed.

In this article, we present a novel approach, *Inforence*, for fault localization using an information-theory based feature selection algorithm. *Inforence* employs a dynamic weighting based feature selection algorithm, inspired from [12-14], which not only selects the most relevant program statements and eliminates redundant ones but also tries to recognize groups of interdependent statements which altogether may affect the program failure. To this aim, relevance, interdependence, and redundancy analysis are performed using information theoretic criteria. Instead of directly using scores computed by a feature selection method to localize faults, *Inforence* employs a method based on statistical causal inference to estimate the failure-causing effect of selected program statements. As a result, unlike existing machine learning based fault localization methods, confounding bias problem [15], which its negative impact on the performance of fault localization has been shown in recent works [15-18], is addressed. More importantly, by performing feature selection and statistical causal inference in a combinatorial manner, we have succeeded to leverage two significant limitations of existing causal inference based methods, their scalability issues due to considerable computational and profile storage overheads and their poor performance in the case of programs containing bugs with combined causes.

*Inforence* also takes advantage of the strength of program slicing [19] in restricting the statements to those included in the cause-effect chain, chains that link bug cause(s) and bug-related statements according to their relations on program dependency graph (PDG), of the failure(s).

In summary, *Inforence* has following contributions:

1. Constructing program spectrum based on program slicing.
2. Employing a novel method based on information theoretic analysis to consider the simultaneous effect of program statements on each other and on the program termination status.
3. Constructing cause-effect chain(s) of program failure using candidate faulty statements selected through information theoretic analysis and ranking them based on a causal inference based method.
4. Evaluation of effectiveness and efficiency of proposed method across various programs. The line of code (LOC) of these programs ranges from 141 to 59,864, peer approaches include most of the representative spectrum-based, including machine learning based, fault localization techniques.

The remaining part of the paper is organized as follows. Section 2 presents a motivating example that illustrates the idea behind this article. The details of proposed method are described in section 3. The experiments and results are shown in section 4. Some discussions containing the related works and threats to validity are presented in sections 5 and 6, respectively. Finally, the concluding remarks are mentioned in section 7.

## 2 Motivating example

The example presented in Table 1 demonstrates the advantage of analyzing the combinatorial effect of program elements on program termination status for localizing faults. As stated before, a large number of program faults involve complex interactions between many program statements (We call these type of bugs as complex bug). These program elements are often related to each other regarding data and control dependencies in PDG. So, after identification of candidate faulty statements, cause-effect chains of failure can be constructed by linking them to each other. In the case of complex bugs, SBFL techniques are not likely to locate the faulty statements accurately without taking this fact into consideration [20]. However, the proposed information-theory based framework applies a dynamic weighting based approach which results in the selection of not only the most relevant program elements (with eliminated redundant elements), but also useful intrinsic groups of interdependent elements. Although, statistical causal inference based methods like [15-18][21-24] attempts to take into consideration multiple statements in another form. Given a statement *s* in program P, they obtained a causal effect estimate of s on the outcome of P that is not subject to severe confounding bias, i.e., a causation-based suspiciousness score of s that takes into consideration other statements that relate to s via control/data dependence. It is important to note that in the case of complex bugs, the execution of faulty statement is not sufficient condition to reveal the failure and the faulty statement is not the only cause of the failure. So, these methods are also incapable of estimating the combinatorial causal effect of multiple statements on program failure, which means that they provide poor performance in the case of complex bugs. Moreover, they have significant limitations and suffer from scalability issues [16].

Consider the program in Table 1, the source code of a small calculator with a seeded semantic fault. The program has three input parameters: two numbers as operands and an integer parameter, between 1 and 3, as the operator. It has an error concerning the assignment in statement S9, i.e., instead of assigning a to *rmax*, we accidentally assign b to *rmax*. We also have a set of twelve test cases out of which six execute successfully (t7, t8, t9, t10, t11, t12), while the other six (t1, t2, t3, t4, t5, t6) result in failure. The coverage information for each test case is also shown where a dot indicates that the corresponding statement is covered, and the absence of a dot indicates that the test case does not cover the statement. In this example, there are two conditional statements in which the result of the program would be incorrect. The first statement is S8, which evaluates whether the first parameter, a, is greater than the second parameter, b. The second conditional statement is S14 which is evaluated as true if the value of the input parameter c is 3. In other words, an incorrect output is generated if and only if both statements S9 and S15 are executed. Therefore, the combination of these two statements causes failure when the program is executed with failing input parameters.

**Table 1** a) Sample program with coverage and execution results for each test case.

| Stmt. #. | Program($P$) | Coverage | | | | | | | | | | | |
|---|---|---|---|---|---|---|---|---|---|---|---|---|---|
| | | T1 a=4 b=1 c=3 | T2 a=7 b=6 c=3 | T3 a=6 b=3 c=3 | T4 a=2 b=1 c=3 | T5 a=3 b=2 c=3 | T6 a=9 b=7 c=3 | T7 a=8 b=-3 c=2 | T8 a=9 b=-2 c=2 | T9 a=-6 b=8 c=3 | T10 a=7 b=6 c=1 | T11 a=6 b=8 c=3 | T12 a=-8 b=9 c=3 |
| $S_1$ | **read**(a,b,c); | * | * | * | * | * | * | * | * | * | * | * | * |
| $S_2$ | result = 0; | * | * | * | * | * | * | * | * | * | * | * | * |
| $S_3$ | rdiv = 1; | * | * | * | * | * | * | * | * | * | * | * | * |
| $S_4$ | rsum = a + b; | * | * | * | * | * | * | * | * | * | * | * | * |
| $S_5$ | **if** ((a > 0) && (b > 0)) | * | * | * | * | * | * | * | * | * | * | * | * |
| $S_6$ | rdiv = a / b; | * | * | * | * | * | * | | | | * | * | |
| $S_7$ | rmax = b; | * | * | * | * | * | * | * | * | * | * | * | * |
| $S_8$ | **if** (a > b) | * | * | * | * | * | * | * | * | * | * | * | * |
| $S_9$ | rmax = b; //Correct: rmax = a; | * | * | * | * | * | * | * | * | | * | | |
| $S_{10}$ | **if** (c == 1) | * | * | * | * | * | * | * | * | * | * | * | * |
| $S_{11}$ | result = rsum; | | | | | | | | | | * | | |
| $S_{12}$ | **if** (c == 2) | * | * | * | * | * | * | * | * | * | * | * | * |
| $S_{13}$ | result = rdiv; | | | | | | | * | * | | | | |
| $S_{14}$ | **if** (c == 3) | * | * | * | * | * | * | * | * | * | * | * | * |
| $S_{15}$ | result = rmax; | * | * | * | * | * | * | | | * | | * | * |
| $S_{16}$ | **return** result | * | * | * | * | * | * | * | * | * | * | * | * |
| Execution Result (0=Successful/ 1=Failed) | | **1** | **1** | **1** | **1** | **1** | **1** | **0** | **0** | **0** | **0** | **0** | **0** |

**Table 1** b) Suspiciousness scores computed using fault localization methods

| Technique name | Formula expression for Suspiciousness score calculation | Ranking of top 3 statements | | |
|---|---|---|---|---|
| **Ochiai[7]** | $\dfrac{N_{CF}(p)}{\sqrt{N_F \times (N_{CS}(p) + N_{CF}(p))}}$ | S6=0.87 | S9=0.81 | S15=0.81 |
| **O[8]** | $\begin{cases} -1 & if\ N_{UF}(p) > 0 \\ N_{US}(p) & otherwise \end{cases}$ | S6=4 | S9=3 | S15=3 |
| **GP19[48]** | $N_{CF}(p) \times \sqrt{|N_{CS}(p) - N_{CF}(p) + N_{UF}(p) - N_{US}(p)|}$ | S6=16.97 | S9=14.69 | S15=14.69 |
| *Inference* | Algorithm 1 presented in page 23 | S9=0.63 | S15=0.56 | S6=0.21 |

Since neither S9 nor S15 on their own has a strong effect on program results, their analysis in isolation cannot help pinpoint the fault, as can be seen in part (b) of Table1. This is because the true observation of S8 in executions in which the value of parameter c is not 3 does not cause failure. In fact, statements S9 and S15 in the motivating example have a grouping effect on program failure.

We can see that the interaction between faulty statement and other correlated statements are likely to cause coincidental correctness (CC), which is a well-known challenge in SBFL [25-27]. To reduce the negative impact of CC tests on fault localization performance and to be able to localize this type of bugs, it is required to analyze the joint impact of statements on the program failure.

*Inforence* performs relevance, interdependence, and redundancy analysis and is capable of determining these statements as joint causes of failure. Considering that S15 has a data dependency to S9 on PDG, *Inforence* will link these two statements to each other and construct cause-effect chain(s) of program failure. The results of applying *Inforence* and three well-known SBFL techniques on our example are presented in part (b) of Table 1. Compared techniques identify S6 as the most fault suspicious statement. They cannot assign the highest score to the faulty statements S9 and S15, because their individual presence in failing runs is not as large as the presence of S6. The existence of CC tests also negatively affects their performance. The results show that our combinatorial analysis yields more accurate results when compared to the three mentioned analysis techniques.

## 3 The method overview

The framework of our approach is shown in Fig. 1.

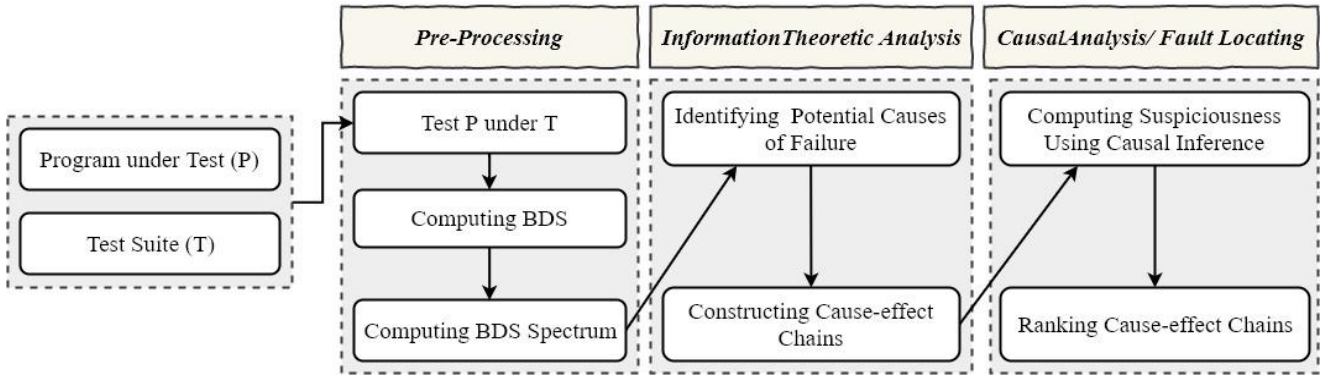

**Fig. 1.** The framework of our approach

Our approach is mainly composed of three modules: pre-processing module, information theoretic analysis module, and causal analysis/fault locating module. The initial inputs are a buggy program and a test suite. To compute the backward dynamic slices (BDS) of failing and passing runs, we use the dynamic slicing framework introduced in [28]. The framework instruments a given program and executes a GCC compiler to generate binaries and collects program dynamic data to produce its dynamic dependence graph. In the following, the building blocks of our framework is introduced in detail.

### 3.1 Preprocessing stage

The first phase of our approach focuses on preprocessing. To minimize the search space in our approach, we suggest the use of BDSs of program executions, instead of all the program statements, to build the spectra. Program spectra, also known as code coverage, represents the set of components covered during test execution. SBFL techniques use information from program entities executed by test cases to indicate entities more likely to be faulty.

In fact, a major difficulty with the current SBFL approaches is to consider all the program features such as statements, branches, and functions to build the program spectrum. This results in large search space which makes it difficult to look for fault predictors. On the other hand, BDS just selects the statements from failure outputs' execution paths and therefore can be considered as *fault candidate statements*. In the following, the process of building program spectra is described. To this end, some basic definitions are depicted first.

**DEFINITION 1.** Given program *P* and a set of test cases

$Tc_P$: each test case $tc_j$ consists of input parameters, $I_j$, and the corresponding desired output, $O_j$. After executing $P$ on $tc_j$, $Exe_P(tc_j)$, the output result would be $O'_j$, $Exe_P(tc_j)= O'_j$. Test case $tc_j$ is called failing test case, $tc_j^{fail}$, if: $O_j \neq O'_j$ and otherwise it is named passing test case, $tc_j^{pass}$. Therefore, the set of test cases in $Tc_P$ is split into two disjoint categories: $Tc_P^{pass}$ for passing test cases and $Tc_P^{fail}$ for failing test cases.

**DEFINITION 2**. Running program $P$ with test case $tc_k$ generates a dynamic dependence graph, $G_P^K(N,E)$, having $N$ nodes and $E$ directed edges: Each node $S_j \in N$ represents the $j^{th}$ execution instance of statement $S$ in the program and an edge $S_j$ to $T_i$ stands for a dynamic data or control dependence from the $j^{th}$ execution instance of statement $S$ to the $i^{th}$ execution instance of statement $T$.

**DEFINITION 3.** Let $G_P^K(N,E)$ be a dynamic dependence graph for the program $P$ running $tc_k$. The BDS of $S_j$, $BDS^k_P(S_j)$, is a sub-graph of $G_P^K(N,E)$ which is constructed by backward traversal of $G_P^K(N,E)$ from $S_j$ to the last reachable node of the graph. The remaining part of $G_P^K(N,E)$ contains nodes and edges which are executed but are not presented in $BDS^k_P(S_j)$. Since failure in a failing run of a given program is mostly manifested as a wrong output value, the BDS of the value frequently captures the faulty code responsible for producing the incorrect value [39]. Therefore, for each failing program run, execution instances corresponding to the erroneous output statements are identified and backward slices of such instances are computed.

The first step in computing BDSs and constructing execution vectors includes analyzing a given program and identifying the output statement(s) to find out which statements generate incorrect results. Some programs may have multiple outputs, and the debugger should determine the one producing incorrect values. The primary concern is how to consider data and control dependence among the output value and the incorrect code when analyzing different failing and passing runs. Since we do not know the location of the faulty code, we compute the backward full dynamic slice starting from the incorrect value which frequently captures the faulty statement(s).

Assume that program $P$ has an output statement, $OP$, with $k$ execution instances, $OP=\{op_1, op_2,..., op_k\}$, where it produces incorrect values for some specific test cases.

Here, for simplicity, we could choose a single execution instance of $OP$ that has an erroneous value in at least one single execution of $P$. For such $op_j$ the BDS is computed for failing test case $tc_m^{fail}$, $BDS^m_P(op_j)$, and passing test case $tc_n^{pass}$, $BDS^n_P(op_j)$. For programs with more than one incorrect output, simply the union of backward slices for all incorrect value outputs is computed. With two categories of failing and passing test cases and the computed backward slices for all existing test cases, the program spectra can be constructed.

**Table 3** Notations used in this paper

| Notation | Description |
| --- | --- |
| P | Program under test |
| $T_{cP}$ | Set of test cases |
| $tc_j$ | A test case in $T_{cP}$ |
| $G_P^k(N,E)$ | Dynamic dependence graph of P when executing with $tc_k$ |
| $BDS_P^k(S_j)$ | Dynamic backward slice of P when executing with $tc_k$ with respect to $S_j$ |
| $I_\phi(X;Y)$ | The mutual information of $X$ and $Y$ |
| $H_\phi(X)$ | The entropy of $X$ |
| $I_\phi(X;Y|Z)$ | The conditional mutual information of $X$ and $Y$ |
| $RR(i,j)$ | Relative Redundancy Ratio of $S_i$ and $S_j$ |
| $IR(i,j)$ | Interdependent Ratio of $S_i$ and $S_j$ |
| $CR(i,j)$ | Correlation ratio of $S_i$ and $S_j$ |
| $N_{CF}(s)$ | Number of failed test cases that cover $s$ |
| $N_{UF}(s)$ | Number of failed test cases that do not cover $s$ |
| $N_{CP}(s)$ | Number of passed test cases that cover $s$ |
| $N_{UP}(s)$ | Number of passed test cases that do not cover $s$ |
| $N_F$ | Total number of passed test cases |
| $N_P$ | Total number of failed test cases |
| $\hat{\tau}(s)$ | Failure-causing effect of $s$ |

### 3.2 Information theoretic analysis stage

In this section, we introduce an information-theory based technique to identify those bug-related statements which may cause the program to fail. Regarding the fact that in most cases multiple statements cause the program to fail in a grouping manner, the technique attempts to keep groups of interdependent statements which altogether may affect the program failure. It is important to note that the combined causes considered in this paper are different from the generally discussed multiple causes. A program failure may have multiple causes due to the existence of multiple bugs in the program. However, by combined causes, we mean that multiple causes collectively cause the program to fail. After identification of likely causes of program failure, using proposed information theoretic analysis, the selected bug-related statements are chained to each other based on the program static structure, to provide the programmer with the context of failure.

### 3.2.1 Preliminaries of information theory

The fundamental concepts of information theory [30], entropy and mutual information, provide intuitive tools to quantify the amount of uncertainty involved in the value of random variables and the information shared by them. Let X be a discrete random variable and probability density function $p(x) = Pr(X = x)$. The entropy $H(X)$ of a discrete random variable $X$ is defined by:

$$H(X) = -\sum_{x \in X} p(x) \log p(x) \quad (1)$$

In [10], authors proposed a family of generalized entropies which is defined using the following functional:

$$H_{\varphi,\phi}(X) = \Psi\left(-\sum_{x \in X} \frac{\phi(p(x))}{\phi'(p(x))}\right) \quad (2)$$

Where $\varphi: (0,1] \to (-\infty, 0)$, is strictly monotonically increasing and is concave with $\varphi(0) \to -\infty$ and $\varphi(1) = 0$. It is clear from the definition if $\varphi(x) = \log(x)$ and $\Psi(x) = x$, $H_{\varphi,\phi}(X)$, reduces to the Shannon entropy Eqn. (1).

They reported that using mutual information that is based on generalized entropies leads to more accurate fault localization. So, we make use of their generalized definition of entropy in our approach.

Joint entropy $H(X, Y)$ extend the definition of entropy $H(X)$ to a pair of random variables and is defined as:

$$H_{\varphi,\phi}(X, Y) = \Psi\left(-\sum_{x \in X, y \in Y} \frac{\phi(p(x,y))}{\phi'(p(x,y))}\right) \quad (3)$$

Conditional entropy $H(X|Y)$ is defined as the entropy of a random variable $X$ conditional on the knowledge of another random variable $Y$. The conditional entropy $H(X|Y)$ is:

$$H_{\varphi,\phi}(X|Y) = \Psi\left(-\sum_{x \in X, y \in Y} \frac{\phi(p(x|y))}{\phi'(p(x|y))}\right) \quad (4)$$

Mutual information (MI) is a measure of the amount of information shared by two variables $X$ and $Y$. Consider two random variables $X$ and $Y$, the mutual information, $I(X; Y)$, is defined as:

$$I(X;Y) = \sum_{x \in X}\sum_{y \in Y} p(x,y) \log \frac{p(x,y)}{p(x)p(y)} \quad (5)$$

Mutual information $I(X; Y)$ can be rewritten as $I(X;Y) = H(X) - H(X|Y)$. Thus, MI can be described as the reduction in the uncertainty of one random variable due to the knowledge of the other.

In terms of information theory, the more relevant statement shares more information with program termination state. Due to difficulties in computing generalized mutual information using generalized entropies in their closed form analytical expressions, we employ a linearized version of mutual information that would approximate the actual mutual information for generalized entropies, introduced by Burbea and Rao [31]. Being linear, this formulation is computationally tractable and is given by following equation:

$$I_\phi(X;Y) = H_\phi(X) - \sum_{y \in Y} p(y) H_\phi(X|Y) \quad (6)$$

Where $H_\phi(X) = -\sum_{s \in X} \phi(p(x))$ and $\phi(x)$ is a convex function.

Conditional mutual information (CMI) is defined as the expected value of the mutual information of two random variables $X$ and $Y$, given the value of a third variable $Z$. It is formally defined by:

$$I_\phi(X;Y|Z) = H_\phi(X|Z) - \sum_{y \in Y} p(y) H_\phi(X|Y,Z) \quad (7)$$

CMI is also used as the reduction in the uncertainty of $X$ due to knowledge of $Y$ when $Z$ is given. Mutual Information $I(X;Y)$ yields values from $0$ to $+\infty$. The higher the $I(X;Y)$, the more information is shared between $X$ and $Y$. However, high values of mutual information might be unintuitive and hard to interpret due to its unbounded range of values. To overcome this issue, mutual information should be normalized. In this paper, the symmetrical measure named symmetric uncertainty (always lies between 0 and 1) is used, given by:

$$U(X,Y) = 2 \frac{I_\phi(X;Y)}{H_\phi(X) + H_\phi(Y)} \quad (0 \leq U(X,Y) \leq 1) \quad (8)$$

Where, $I_\phi(X;Y)$ is the mutual information and measures the amount of information shared by two variables $X$ and $Y$. $H_\phi(X)$ and $H_\phi(Y)$ also represent the entropy of variables $X$ and $Y$, respectively.

So, the relevance of feature f for the target class can be denoted as:

$$R(f, Out) = U(f, Out) \quad (9)$$

3.2.2 Constructing cause-effect chains of failure using information theoretic analysis

Conventional feature selection algorithms tend to select features which have high relevance with the target class and low redundancy among the selected features. The major disadvantage of these algorithms is that they ignore the dependencies between the candidate feature and unselected features. In most previous literature, candidate features which are highly correlated with the selected features will be discarded. However, it is likely to ignore features which as a group have strong discriminatory power but are weak as individuals. For example, in fault localization context, program statements normally function in groups and have a combinatorial impact on program termination status. Statements in such a groups are highly correlated, and each statement cannot function apart from one another. Therefore, the traditional criteria are unsuitable for such applications. As stated before, it is required to estimate the simultaneous impact of different combinations of program features (e.g. statements) on the program failure, dependent on the appearance of the statements in cause-effect chains of various executions of the program. To tackle this problem, we introduce a new scheme that can evaluate the redundancy and interdependence of program statements. Now, we describe the relevance, interdependence and redundancy analysis performed using information theoretic criteria.

**DEFINITION 4. (Redundant feature)** A feature (i.e. program statement) is said to be redundant if one or more of the other features are highly correlated with it and its relevance with the program termination status can be reduced by the knowledge of any one of these features.

$$I_\phi(s_i; Out|s_j) < I_\phi(s_i; Out) \quad (10)$$

$I_\phi(s_i; Out|s_j)$, represents the conditional mutual information of $s_i$ and $Out$ given $s_j$ and indicates the reduction in the uncertainty of $s_i$ due to the knowledge of $Out$ when $s_j$ is given.

**DEFINITION 5. (Relative Redundancy Ratio)** $RR(i,j)$ represents the ratio of the reduction of relevance between statement $s_i$ and $out$ (i.e. program termination status) due to the statement $s_j$. A value of 0 indicates that the relevance between $s_i$ and $out$ is not reduced by the knowledge of $s_j$.

$$RR(i,j) = 2\frac{I_\phi(s_i; Out|s_j) - I_\phi(s_i; Out)}{H_\phi(s_i) + H_\phi(Out)} \quad (11)$$

where $(-1 \leq RR(i,j) \leq 0)$

**DEFINITION 6. (Interdependent Ratio)** Two statements $s_i$ and $s_j$ are interdependent on each other if the following form is satisfied.

$$I_\phi(s_i; Out|s_j) \geq I_\phi(s_i; Out) \quad (12)$$

$IR(i,j)$ denotes the ratio of the increase of relevance between statement $s_i$ and $out$ by new statement $s_j$ joining.

$$IR(i,j) = 2\frac{I_\phi(s_i; Out|s_j) - I_\phi(s_i; Out)}{H_\phi(s_i) + H_\phi(Out)} \quad (13)$$

where $(0 \leq IR(i,j) \leq 1)$

Both $RR(i,j)$ and $IR(i,j)$ are unified as correlation ratio $CR(i,j)$, is given as formula (7), which will be used in the algorithm presented in the following section. Correlation ratio restricts its values to the range [-1,1].

$$CR(i,j) = \begin{cases} IR(i,j) & if: I_\phi(s_i; Out|s_j) > I_\phi(s_i; Out) \\ RR(i,j) & if: I_\phi(s_i; Out|s_j) \leq I_\phi(s_i; Out) \end{cases} \quad (14)$$

Now we present the *Cause-Effect Chains Construction* algorithm. At first, program statements get an initial weight (these weights could be adjusted according to the fault-proneness of different parts of the program). Then the weighted statements are sent to the feature evaluation module, and one candidate statement is selected by feature evaluating. The weight of each statement reflects its impact of correlation (interdependence and redundancy) on the already selected subset. Accordingly, the weight of the rest candidate statements will be dynamically updated according to their correlation with the newly selected statement. This process will be repeated until all statements are ranked or a limitation prevails.

The parameter δ is a user-specified threshold to terminate the procedure. Since, the majority of faults in subject programs are localizable with 20-30% of manual code inspection in the worst case, we set the threshold value, δ, to **30% × |PS|**, where PS represents the *potentially faulty candidate statements*. If we reduce the value of the threshold, the computational cost can be reduced but some faults may not be localizable.

In the beginning (lines 1–3), it initializes relative parameters, weights each statement equally (or based on its fault-proneness likelihood) and calculates the value of $R(s, Out)$ for every statement. Then it enters the iteration and does not leave it until select more than $\delta$ statements. In each iteration, it firstly calculates the value of criterion $J(s)$ for each statement (lines 5–7). The statement s with the largest $J(s)$ will be chosen and moved from the potentially faulty statements set $PS$ to subset $S$ (lines 8–10). After selecting bug-related statements in each iteration, we check how these statements relate to each other in PDG and construct cause-effect chains of failure.

Determining cause-effect chains allows a programmer to focus his attention on the most relevant part of the program to the failure. In fact, we exploit the transitional information provided by data/control dependencies among program statements to expose more bugs and discover cause-effect chains.

In order to select the optimal statement in this stage, evaluation criterion $J(s)$ is defined to evaluate the superiority of each statement over others. As mentioned above, $R(s, Out)$ indicates the inherent relevance of statement $s$ with program failure and $w(s)$ denotes its impact on the already selected statements. Therefore, we utilize the weight $w(s)$ to regulate the relative importance of the relevance $R(s, Out)$ considering its interdependent and redundant relationship with the selected statements. Finally, the weight of the rest candidate statements will be updated (lines 11–14). The variation of the weight for each statement $s_j$ is determined by its correlation ratio $-1 \leq CR(s_j, s) \leq 1$ with the newly selected statement $s$. The reweighted statements are really wanted in the next iteration. The selection procedure will be terminated if the number of selected statements is larger than the user-specified threshold $\delta$. It is worth to mention that existing feature selection based fault localization algorithms determine discriminative program elements based on their correlation with program termination status. They identify program elements that can differentiate instances of passed and failed executions. Unfortunately, they do not differentiate program elements that characterize failed executions from those that characterize passed executions [32]. Two program elements can both have high discriminative power but characterize the different class of executions. If a program element characterizes the failed executions, a higher score indicates a higher likelihood for the corresponding program element to be faulty. However, if a program element characterizes the passing executions, a higher score indicates a lower likelihood for the corresponding program element to be faulty. Therefore, program elements that identified as discriminative and important by a standard feature selection method may not necessarily be related to fault locations. To overcome this issue, as mentioned in [32], there is a need to separate important and discriminative program elements based on how close they are to a class of executions. To this aim, we use the following function:

$$RC(s) = \begin{cases} +1 & if \ \frac{N_{CF}(s)}{N_F} - \frac{N_{UF}(s)}{N_F} > \frac{N_{CP}(s)}{N_P} - \frac{N_{UP}(s)}{N_P} \\ -1 & if \ \frac{N_{CF}(s)}{N_F} - \frac{N_{UF}(s)}{N_F} \leq \frac{N_{CP}(s)}{N_P} - \frac{N_{UP}(s)}{N_P} \end{cases} \quad (15)$$

$RC(s)$ is defined based on the notations of SBFL statistics to determine which class (i.e., failed or passed) that s is closer to. $N_{CF}(s)/N_F - N_{UF}(s)/N_F$ is the ratio between the number of failed test cases that execute s and the total number of failed test cases, considering the noise in fault-failure correlation measurements that introduced by other statements on the same set of executions that may lead to the observed failures [33]. A higher value of this ratio indicates a stronger relationship between s and failed test cases. $N_{CP}(s)/N_P - N_{UP}(s)/N_P$ is the ratio between the number of passed test cases that execute s and the total number of passed test cases, considering the noise in fault-failure correlation measurements. A higher value of this ratio indicates a stronger relationship between s and passed test cases.

Step-by-step calculations of our feature selection method and the resultant cause-effect chains for motivating example are shown in Table 2.

**Algorithm** *Cause-Effect Chains Construction*

**Inputs**
  1) Static program dependence graph: S-PDG = (V, E)   2) Test execution set: Tests = {t1, t2, ... , tn}
  3) Slice coverage spectrum: SCP = [tj, s]   4) Potentially faulty candidate statements: PS

**Outputs**
  Cause-effect chains for program faults

(1) Initialize parameters: $k = 0$, $S = \phi$;
(2) Initialize the weight $w(s)$ for each statement $s$ in $PS$ to $w(s) = 1 +$ fault proneness likelihood(s)
(3) Calculate the value of $R(s, out)$ for each statement $s$ in $PS$;
(4) **While** $k \leq \delta$ **do**
(5)   **For** each candidate statement $s \in PS$ **do**
(6)     Calculate the value $J(s) = R(s, out) \times w(s) \times RC(s)$;
(7)   **End**
(8)   Choose the candidate statement $s_a$ with the largest $J(s)$;
(9)   Add $s_a$ into the selected subset S, i.e., $S = S \cup \{s_a\}$;
(10)  $PS = PS - \{s_j\}$;
(11)  **For** each candidate statement $i \in (PS \cup S)$ **do**
(12)    Calculate the $CR(i, j)$;
(13)    Update $w(i)$ by $w(i) = w(i) \times (1 + CR(i, j))$;
(14)  **End**
(15)  k = k + 1

//Generating Cause-effect chains
 1- Select the next statement, $s_b$, according to the lines 5-8
 2- Check the adjacency matrix of S-PDG to find whether the statement corresponding to $s_b$ has direct data/control dependence with the statement corresponding to statement $s_a$.
   a. If there is direct dependence, put $s_b$ in the chain containing $s_a$ and according to the direction of the dependence edge in the static graph, link the statements $s_b$ to $s_a$.
   b. If there is no direct dependence, and the number of chains is no more than $k'$, make a new chain containing $s_b$.
 //In the case of the next selected statements such as $s_c$:
 3- Check the static dependence graph to find whether the statement corresponding to $s_c$ has direct data/control dependence with the statements corresponding to statements $s_a$ or $s_b$.
   a. If it has direct dependence with both $s_a$ and $s_b$, merge two chains containing $s_a$ and $s_b$ and add $s_c$ to the newly merged chain.
   b. If it has direct dependence with either $s_a$ or $s_b$, put $s_c$ in the chain containing the dependent statement and according to the direction of the dependence edge in the static graph, link the statement $s_c$ to the dependent variable.
   c. If there is no direct dependence, and the number of chains is no more than $k'$, make a new chain containing $s_c$.
(16) **End**

**Algorithm. 1**. Cause-effect chains construction algorithm

Due to space limitations, calculations are provided for some of the statements, including those that have a grouping effect on program failure. In the first iteration of the Algorithm, the relevance of statements to program termination state is measured using equation 9 and accordingly the value of $J_1(s)$ is calculated. Since the relevance analysis is performed for each statement in isolation, s6, similar to other SBFL techniques, gains the highest weight. So, s6 is selected as the first statement and the chain1 is constructed. The weight of each statement reflects its impact of correlation (interdependence and redundancy) on the already selected subset. Accordingly, the weight of the rest candidate statements is dynamically updated (line 13) according to their correlation with the newly selected statement, s6. In the second iteration, s9 and s15 jointly gain the highest weight. So, one of them (assume s9) is selected. Again, the weight of the rest candidate statements (including the already selected statements) is dynamically updated according to their correlation with s9. We can see that in the third step, the weight of s6, one of previously selected statements, is considerably reduced and s15 gains the highest weight. So, s15 is selected and linked to s9 in chain2.

**Table 2** Step-by-step calculations of our feature selection method (algorithm 1) and resultant cause-effect chains

| measures | Candidate Statements | | | | |
|---|---|---|---|---|---|
| | S6 | S9 | S11 | S13 | S15 |
| *First Iteration* | | | | | |
| $R(s_i, out)$ | 0.48 | 0.34 | 0.12 | 0.23 | 0.34 |
| $RC(s_i)$ | 1 | 1 | -1 | -1 | 1 |
| $W(s)$ | 1 | 1 | 1 | 1 | 1 |
| $J_1(s)$ | 0.48 | 0.34 | -0.12 | -0.23 | 0.34 |
| *Second Iteration* | | | | | |
| $CR(s_6, s_{i,i\neq 6})$ | | -0.12 | 0.15 | -0.23 | -0.12 |
| $W(s_{i,i\neq 6})$ | | 0.88 | 1.15 | 0.77 | 0.88 |
| $J_2(s)$ | | 0.30 | -0.14 | -0.18 | 0.30 |
| *Third Iteration* | | | | | |
| $CR(s_9, s_{i,i\neq 9})$ | -0.12 | | 0.08 | 0.18 | 0.41 |
| $W(s_{i,i\neq 9})$ | 0.88 | | 1.24 | 0.91 | 1.24 |
| $J_3(s)$ | -0.10 | | -0.15 | -0.21 | 0.42 |

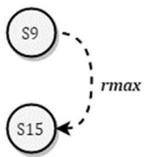

Chain1 with one statement    Chain2 with two statements

It is important to note that the concept of cause-effect chain of failure is first presented in Delta debugging [34-35]. It isolates the relevant variables and values by systematically narrowing the state difference between a passing run and a failing run—by assessing the outcome of altered executions to determine whether a change in the program state makes a difference in the test outcome. Zellers' cause-effect chain consists of the variables and values that caused the failure.

### 3.3 Statistical causal inference of failure cause-effect chains

As described above, along with the information theoretic analysis and identification of candidate faulty statements, the cause-effect chains of failure are constructed.

According to identified candidate faulty statements and existing data/control dependencies amongst them in PDG, the following two cases may occur:

1- Several cause-effect chains are constructed. This is because of two reasons. First, when linking the identified candidate faulty statements to each other based on PDG, a unique cause-effect chain may not be formed because the likely causes of failure do not explain the failure context perfectly and thus several cause-effect chains are constructed. Second, in the case of programs containing multiple faults, several cause-effect chains may be created that each of which may correspond to one of the several faults in a program.
2- The resultant cause-effect chain(s) may consists of many statements in larger programs and thus locating root cause(s) of failure could be difficult for the programmer.

In order to deal with the difficulties arising in these two cases, we employ a statistical causal inference method, based on existing works in the literature [15-18][21-24], to estimate the causal effect of statements incorporated in resultant cause-effect chain(s) on program failure. We then prioritize resultant chains based on estimated causal effects of statements included in each of them.

Recent SBFL studies [15-18][21-24] has demonstrated that the suspiciousness score of a program statement can be improved by statistically controlling for a small number of potential confounding factors involving the statement's direct program-dependence predecessors.

Baah et al. [15-17] showed that Pearl's theory of causal graphical models [36], in particular his well-known Backdoor Criterion, could be best applied in SBFL to a causal graph (causal DAG) derived from a PDG [37-38] with the aim of mitigating confounding bias when estimating the causal effect of executing a statement on the occurrence of program failures. A causal diagram is a graphical representation of causal relationships among variables [26]. Causal diagrams represent assumptions about the causal relationships that are relevant to a problem, and hence they guide the process of making causal inferences from data.

Since Baah et al.'s techniques [15-17] involve fitting a regression model or computing a matching for each statement in a program, their computational and profile storage overheads may be considerable for large programs. Therefore, they suffer from scalability issues. One way to reduce the costs of collecting, storing, and analyzing execution profiles is to assign suspiciousness scores to program units or regions of larger size than individual statements [39]. However, in the case of large program units, the programmer may still have a difficult task in locating the fault and a large amount of manual code inspection may be required. There is a valid implicit assumption that software developers can consistently recognize faults by examining only "suspicious" program statements. Therefore, another way to reduce the costs is to make causal inference on a small subset of program statements that have been identified as candidate faulty statements. We adopt this approach and make causal inference on only statements that included in the resultant cause-effect chain(s) on program failure. On the other hand, existing causal inference based methods are unable to provide good performance in the case of programs containing complex bugs (i.e. bugs with combined causes). Since *Inforence* is able to discover combined causes of program failure through information theoretic analysis described in section 3-2 and form cause-effect chain(s) of failure by linking them to each other based on PDG, we can provide the programmer with more useful information about combined causes of such complex bugs by making causal inference on identified likely causes.

A remarkable factor that a causal inference based method need to consider is the composition of the test suite used to estimate failure-causing effects of program elements. The lack of balance problem occurs if the distributions of confounding variables differ for the treatment and control groups. Assuming that we are estimating the causal effect of executing a statement s on the occurrence of program failure, confounding bias due to dynamic data and control dependences affecting s can be reduced by considering two groups of program executions: those that cover s and those that do not cover s. these two groups are referred to as the treatment and control groups. So, the lack-of-balance problem occurs for s if the control flow paths induced by the tests that cover s differ too much from those induced by the tests that do not cover s.

Lack-of-balance is mitigated in randomized controlled experiments by the random assignment of subjects to the treatment group and to the control group [36]. In practice, the set of test cases used in SBFL is usually given, and the subset of them that cover and do not cover a particular statement s cannot be assumed to be balanced. Matching technique attempts to mimic randomization by identifying a set of program executions that cover a particular statement s that is comparable on all observed confounding variables to a set of program executions that do not cover s. To achieve a relative balance between the treatment and control groups and thus reduce confounding bias, we employ the important causal inference technique propensity-score matching [40] in this paper. In the following, we propose our method to compute failure-causing effects of program statements included in cause-effect chains.

### 3.3.1 Identification of confounding variables

Baah et al. [15-17] showed that Pearl's Backdoor Criterion [36], could be employed in SBFL to a causal directed acyclic graph (causal DAG) derived from a PDG [37-38] with the aim of identifying admissible sets of factors as confounding variables. The intuition behind the back-door criterion is as follows. The back-door paths in the DAG carry spurious associations from X to Y, while the paths directed along the arrows from X to Y carry causative associations. Blocking the back-door paths, by leveraging the elements of a set S that satisfy the back-door criterion, ensures that the measured association between X and Y is purely causative and indicates the causal effect of X on Y [36]. Similar to existing causal inference based approaches [15-18][21-24], we consider direct program-dependence predecessors of a given statement s as its confounders. Given statement s, we form a vector $X_s$ of confounding variables as follows:

$$X_s = (CD_1, \dots CD_k; DD_1, \dots, DD_k) \qquad (15)$$

For a given execution, the confounding variable $CD_i$ (or $DD_i$) takes on the value 1 if and only if the corresponding statement is covered.

### 3.3.2 Matching executions and estimating casual effects

Matching is an increasingly popular method for preprocessing data to improve causal inferences in observational data. It brings some of the benefits of randomized controlled experiments, in terms of reduced confounding bias, to observational studies.

The goal of matching is to reduce the imbalance in the empirical distribution of the confounding variables between the treated and control groups. It involves, for every treatment unit, finding one or more control units that are similar to it, in such a way that balance is achieved between the resulting treatment group and control group with respect to confounding variables.

An important approach to achieving balance in observational studies is the use of propensity scores [40]. For a binary treatment variable T and a vector X of confounding variables, the propensity score $Pr[T = 1|X = x]$ is conditional probability of treatment T=1 given that observed value of X is x. Propensity-score matching involves forming matched sets of treatment and control units that share a similar value of the propensity score. The smaller the difference between the scores, the more similar they are, so there is a pair matching. Therefore, from the original test suite, just a part of it will be kept for the final analysis.

True propensity scores are generally unknown, but they can be estimated from the observations. Logistic regression [41] is the most commonly used method for estimating them in practice. We first fit a distinct propensity score model $M_S$ for each statement $s$, included in resultant cause-effect chains, using the coverage indicator for $s$ as the treatment variable $T_S$ (the response variable in the model) and the elements of the vector $X_S$ of confounding variables, as predictor variables. The model $M_S$ is defined by the log-odds formula:

$$log \frac{Pr[T_s = 1|X_s]}{Pr[T_s = 0|X_s]} = X'_s \beta \qquad (16)$$

The fitted model is then used to predict the treatment propensity $Pr[T_s = 1|X_s = x_s]$ for each test execution. Now, the matching process can be carried out based on estimated propensity scores. The number of matches in matched group affects the bias, precision, and efficiency of matching estimators. With 1:1 matching, a treatment unit is matched to exactly one control unit. This minimizes confounding bias since only the most similar unit is used for matching. In many-to-one (M:l) matching, M control units are matched to each treatment unit. This strategy reduces estimation variance but is likely to increase bias since due to the greater average dissimilarity between matched units. We matched executions using optimal full matching [42] and unmatched executions were discarded.

---

**Algorithm** *failure-causing-effect (DPDG, Tests, SCT, s)*

**Inputs**
1) Dynamic program dependence graph: D-PDG = $(V, E)$
2) Test execution set: $Tests = \{t1, t2, \dots, tn\}$
3) Statement coverage table: $SCT = [tj, s]$
4) Statement: $s \in S_{information-theory}$

**Outputs**
Failure-causing-effect of statement s
1. effect = -1.0;
// Getting matched data using propensity score matching: Lines 2-8;
2. Model(s) = Compute causal model of s
3. Pred(s) = Compute predecessors of s from Model(s)
//Fit a propensity score model $M_s$ using Equation 9.
4. $M_s$ = Logistic_R($SCT[\cdot, s]$, SCT[, Pred(s)]);
//Estimate the propensity score [$tj$] for each test execution $tj$.
5. **for each** test $tj \in Tests$ do
6.    PS[$tj$] = predict($M_s$,SCT[$tj$, s]);
7. **end for**
//Match executions by propensity scores $PS$.
8. $Dmatch$=match($SCT[, s]$, $PS$);
9. Impute potential outcomes from $Dmatch$ using Equation 17;
10. Compute $\hat{\tau}(s)$ using Equation 18.
11. effect =$\hat{\tau}(s)$;
12. **return** effect;

---

**Algorithm. 2**. failure-causing-effect estimation algorithm

Let $M$ be the number of matches for each unit and $J_M(i) = \{j: \text{unit } j \text{ is matched with unit } i\}$ for $i = 1, \dots, N$. For each unit $i$, one of the two potential outcomes is observed, namely $Y_{i1}$ if $T_i = 1$ and $Y_{i0}$ if $T_i = 0$. The missing potential outcome for unit $i$ is imputed using the average of the outcomes for its matches:

$$\hat{Y}_{i0} = \begin{cases} \frac{1}{M}\sum_{j \in J_M(i)} Y_j & if \ T_i = 1 \\ Y_i & if \ T_i = 0 \end{cases} \qquad (17)$$

and

$$\hat{Y}_{i1} = \begin{cases} Y_i & \text{if } T_i = 1 \\ \frac{1}{M}\sum_{j \in J_M(i)} Y_j & \text{if } T_i = 0 \end{cases}$$

Having observed and imputed potential outcomes for each unit, we can estimate the failure-causing effect $\tau$ using the following estimator:

$$\hat{\tau} = \frac{1}{N}\sum_{i=1}^{N}(\hat{Y}_{i1} - \hat{Y}_{i0}) \qquad (18)$$

The process of estimating the causal effect of individual program elements is presented in the algorithm 2.

## 4 Empirical study

In this section, the effectiveness of *Inforence* is empirically evaluated. To this end, we compare our method with some well-known techniques in the context of software fault localization. To show the performance of *Inforence*, the following case studies are designed to evaluate the proposed method in different ways:

1- The performance of *Inforence* in finding the origin of failure is measured according to some well-known evaluation frameworks and compared to state-of-the-art techniques. We evaluate *Inforence* and other techniques on well-known suites described in the next sub-section. The scalability of *Inforence* to find bug(s) in large programs is also considered in this study.

2- In order to investigate how considering grouping effect of *Inforence* could give the capability of finding multiple bugs to our approach, we combine several versions of benchmark programs to generate multiple-bug versions. After conducting experiments, results are compared to other fault localization techniques.

3- The capacity of *Inforence* in finding the context of failure in terms of cause-effect chains has been studied in case study 3.

All experiments in this section are carried out on a 2.66 GHz Intel Core 2 Quad Processor PC with 6 GB RAM, running UBUNTU 9.10 Desktop (i386). To compute BDSs we used the dynamic slicing framework introduced in [28]. The framework instruments a given program and executes a GCC compiler to generate binaries and collects program dynamic data to produce its dynamic dependence graph. The framework contains two main tools: Valgrind and Diablo. The instrumentation is done by Valgrind memory debugger and profiler who also identifies the data dependence among statement execution instances. The Diablo tool is capable of producing control flow graph from the generated binaries. We also used WET tool to compute BDSs from an incorrect output value. We implemented required statistical procedures using the statistical computing environment R and used some R packages such as Glmnet [43] and MatchIt [44].

In our experiments we use seven different sets of programs, the Siemens suite, gzip, grep, sed, space, make, and Bash. These sets correspond to a total of 13 different subject programs. Table 4 gives the number of faulty versions and test cases of each program and the size in terms of lines of code. The seven programs of the Siemens suite have been employed by many fault localization studies [1-11]. The correct versions, 132 faulty versions of the programs and all the test cases are downloaded from The Siemens Suite [45]. As shown in table 5, some faulty versions are eliminated due to the segmentation faults, no failing test case, and existing a bug in a header file, specified in row 5 to 8 of the table, respectively.

**Table 4** Summary of subject programs

| Program | Number of faulty versions | Lines of code | Number of test cases |
|---|---|---|---|
| Print tokens | 7 | 565 | 4130 |
| Print tokens2 | 10 | 510 | 4115 |
| Replace | 32 | 512 | 5542 |
| Schedule | 9 | 412 | 2650 |
| Schedule2 | 10 | 307 | 2710 |
| Tcas | 41 | 173 | 1608 |
| Tot info | 23 | 440 | 1054 |
| Gzip | 23 | 6753~6K | 211 |
| Grep | 18 | 12653~12K | 470 |
| Sed | 20 | 12062~12K | 360 |
| Space | 38 | 6218~6K | 13585 |
| Make | 28 | 20014~20K | 793 |
| Bash | 6 | 59864~59K | 1168 |

Version 1.1.2 of the gzip program (which reduces the size of named files) was downloaded from [45]. Also found at [45] were versions 2.2 of the grep program (which searches for a pattern in a file), 3.76.1 of the make program (which manages building of executables and other products from source code), and 2.0 of the sed program (which parses textual input and applies user-specified transformation to it).

The correct version of the space program, the 38 faulty versions, and a suite of 13585 test cases used in this study are downloaded from [45]. Three faulty versions are not used in our study because none of the test cases fails on these faulty versions.

Our study includes programs which vary dramatically in terms of size, functionality, number of faulty versions and number of test cases. The programs of the Siemens suites are small (less than 1000 lines of code), the gzip and space programs are medium-sized (between 1000 and 10,000 lines of code), the grep, sed and make programs are large (between 10,000 and 20,000 lines of code), while the bash program is very large with more than 59,000 lines of code. This allows us to better generalize the findings and results of this paper.

4.1 Evaluation metrics

We would like to know whether the program lines reported as faulty by *Inforence* technique are the exact origin of failure. If not, the amount of code to be examined manually by the programmer is under question. In this regard, the amount of manual code examination is an important criterion to evaluate the performance of fault localization algorithms. *EXAM* score is a measure that gives the percentage of statements that need to be examined until the first faulty statement is reached. According to [2] the main objective is to provide a good starting point for programmers to begin fixing a bug, rather than identifying all the code that would need to be corrected. Many fault localization studies applied this metric to evaluate the performance of their fault-localization technique [2-6].

In short, the effectiveness of various fault localization techniques can be compared based on the *EXAM* score, and for any faulty program, if the *EXAM* score assigned by technique is less than that of technique, then is considered to be more effective.

*Average number of statements examined* is another metric that gives the average number of statements that need to be examined with respect to a faulty version of a subject program to find the bug. Cause-effect chains are provided to programmer in the form of a sequence of statements. To evaluate the performance of *Inforence* in this case, precision, recall and F-measure metrics are used.

**Table 5** Additional data about faulty versions of Siemens suite used in our experiments

| Programs | Schedule | Schedule2 | Print tokens | Print tokens2 | Replace | Tot info | Tcas | Total |
|---|---|---|---|---|---|---|---|---|
| Versions with segmentation fault | 0 | 1 | 0 | 1 | 2 | 0 | 0 | 4 |
| Versions with no faulty test cases | 4 | 1 | 0 | 0 | 3 | 0 | 0 | 8 |
| Versions with header file errors | 0 | 0 | 2 | 0 | 0 | 0 | 0 | 2 |
| Used version | 5 | 8 | 5 | 9 | 27 | 23 | 41 | 118 |

4.2 Fault localization techniques used in comparisons

We compare our approach with both SBFL and feature selection based fault localization techniques, using the above mentioned evaluation criteria. In our study, following techniques are considered for comparison: D-star [2], with star value of two, Crosstab [3], H3b-H3c [4], Ochiai [7], O-O$^P$ [8], Relief [9], Gen-Entropy [10] and Baah's causal methods [15-16] as well as the theoretically best genetic programming (GP) based fault localization techniques namely GP02, GP03, and GP19 [48]. Results on programs with a single bug and multiple bugs are presented in sections 4-3-1 and 4-3-2, respectively. Description of peer techniques is provided in related works section of the paper.

4.3 Results

It should be noticed that, for *Inforence* and all other techniques used in experiments, the faulty statement may share the same suspiciousness value with several correct statements. As a result, all these statements will be tied for the same position in the ranking. So, in the best case the user need to examine only one of these statements and in the worst case all these statements need to be examined to locate the fault.

In all our experiments we assume both best case and worst case effectiveness. The evaluation results for the programs with single and multiple bugs are presented in sections 4-3-1 and 4-3-2, respectively.

4.3.1 Results on programs with a single bug

In this case study, each faulty version under consideration has exactly one bug, potentially involving multiple statements at different locations in the code or different functions. Many other studies, such as [2-6], take a similar approach. However, our proposed technique is also capable of handling programs with multiple bugs, which is further discussed in section 4-4-2. Table 6 and table 7 present the average number of statements that need to be examined by each fault localization technique across each of the subject programs under study, for both best and worst cases. For example, the average number of statements examined by *Inforence* with respect to the all faulty versions of gzip is 45.2 in the best case and 117.52 in the worst. Again, with respect to the gzip program, we find that the second best technique is O, which can locate all the faults by requiring the examination of no more than 49.57 statements in the best case, and 121.38 in the worst. For D-Star, the best is 54.82 and the worst 119.89. A similar observation also applies to other programs. In contrast, *Inforence,* in the worst case, detects the faults in space by examining 65.41 statements in average whereas O and $O^P$ do so in 62.89 and 64.34 statements, respectively.

**Table 6**   Average number of statements examined with respect to each faulty version (Best case)

|  | Siemens | gzip | grep | sed | space | make | bash |
|---|---|---|---|---|---|---|---|
| *Inforence* | **7.21** | **45.2** | **90.94** | **70.32** | **29.9** | **119.45** | 69.33 |
| D-Star | 13.66 | 54.82 | 148.21 | 89.38 | 39.28 | 271.62 | 82.46 |
| H3b | 11.55 | 70.48 | 165.06 | 145.25 | 52.10 | 304.39 | 118.50 |
| H3c | 11.20 | 70.48 | 145.63 | 142.10 | 50.95 | 246.79 | 112.63 |
| Crosstab | 16.07 | 62.19 | 213.75 | 103.40 | 47.16 | 336.00 | 92.25 |
| Ochiai | 14.37 | 60.10 | 164.50 | 91.45 | 41.25 | 275.11 | **64.67** |
| O | 9.90 | 49.57 | 125.06 | 76.45 | 34.08 | 189.36 | 81.89 |
| $O^P$ | 10.02 | 49.57 | 125.06 | 76.45 | 35.50 | 189.36 | 81.89 |
| Relief | 12.51 | 56.46 | 189.84 | 78.21 | 40.25 | 166.33 | 72.90 |
| Gen-Entropy | 9.45 | 50.22 | 118.62 | 79.48 | 37.75 | 159.45 | 74.82 |
| GP02 | 15.2 | 65.5 | 176.65 | 98.72 | 44.62 | 274.35 | 79.12 |
| GP03 | 14.16 | 58.21 | 158.40 | 91.25 | 40.86 | 244.5 | 68.46 |
| GP19 | 14.9 | 62.84 | 166.92 | 90.74 | 45.33 | 259.8 | 78.45 |
| Baah, 2010 | 16.55 | 78.48 | 195.06 | 145.25 | 52.31 | 152.25 | 96.63 |
| Baah, 2011 | 10.82 | 51.48 | 135.63 | 118.10 | 45.85 | 141.28 | 85.44 |

**Table 7**   Average number of statements examined with respect to each faulty version (Worst case)

|  | Siemens | gzip | grep | sed | space | make | Bash |
|---|---|---|---|---|---|---|---|
| *Inforence* | **15.85** | **117.52** | **181.45** | 202.26 | 65.41 | **262.95** | **106.75** |
| D-Star | 20.69 | 119.89 | 225.16 | 196.51 | 72.81 | 452.78 | 117.62 |
| H3b | 18.80 | 139.33 | 255.63 | 251.85 | 97.08 | 483.32 | 196.82 |
| H3c | 18.46 | 139.29 | 236.19 | 248.70 | 96.56 | 425.71 | 185.45 |
| Crosstab | 23.09 | 131.00 | 410.88 | 210.00 | 84.26 | 514.93 | 147.32 |
| Ochiai | 21.62 | 128.90 | 255.06 | 206.05 | 72.89 | 454.04 | 121.92 |
| O | 29.72 | 121.38 | 215.63 | **183.05** | **62.89** | 368.29 | 109.81 |
| $O^P$ | 17.29 | 121.38 | 215.63 | **183.05** | 64.34 | 368.29 | 109.81 |
| Relief | 18.92 | 133.56 | 196.45 | 218.66 | 70.25 | 346.85 | 129.76 |
| Gen-Entropy | 19.54 | 129.44 | 206.61 | 211.79 | 72.28 | 310.33 | 112.60 |
| GP02 | 24.85 | 136.52 | 264.76 | 235.78 | 81.26 | 412.46 | 136.64 |
| GP03 | 19.74 | 124.66 | 244.56 | 211.7 | 70.28 | 388.62 | 118.82 |
| GP19 | 23.62 | 129.81 | 266.35 | 231.6 | 74.63 | 401.5 | 129.75 |
| Baah, 2010 | 29.80 | 169.33 | 275.63 | 231.85 | 29.80 | 312.96 | 169.55 |
| Baah, 2011 | 18.46 | 139.29 | 236.19 | 190.70 | 18.46 | 295.38 | 140.25 |

We now present the evaluation of *Inforence* with respect to the EXAM score. Because of space limitations we only provide figures for four programs: Siemens, space, gzip and sed. Figures for the other 3 sets of programs (grep, make, and bash) were not included here. However, the conclusions drawn with regards to the first four programs are also applicable to the remaining three. Fig. 2 illustrates the *EXAM* score of *Inforence* and other peer techniques on Siemens suite using six subplots. The x-axis represents the percentage of code (statements) examined while the y-axis represents the number of faulty versions where faults are located by the examination of an amount of code less than or equal to the corresponding value on the x-axis. For example, referring to Part (b) of Fig. 2, we find that by examining 10 % of the code *Inforence* can locate 77 % of the faults in the Siemens suite in the best cases and 64 % in the worst, whereas D-Star has 64 % (best) and 52 % (worst).

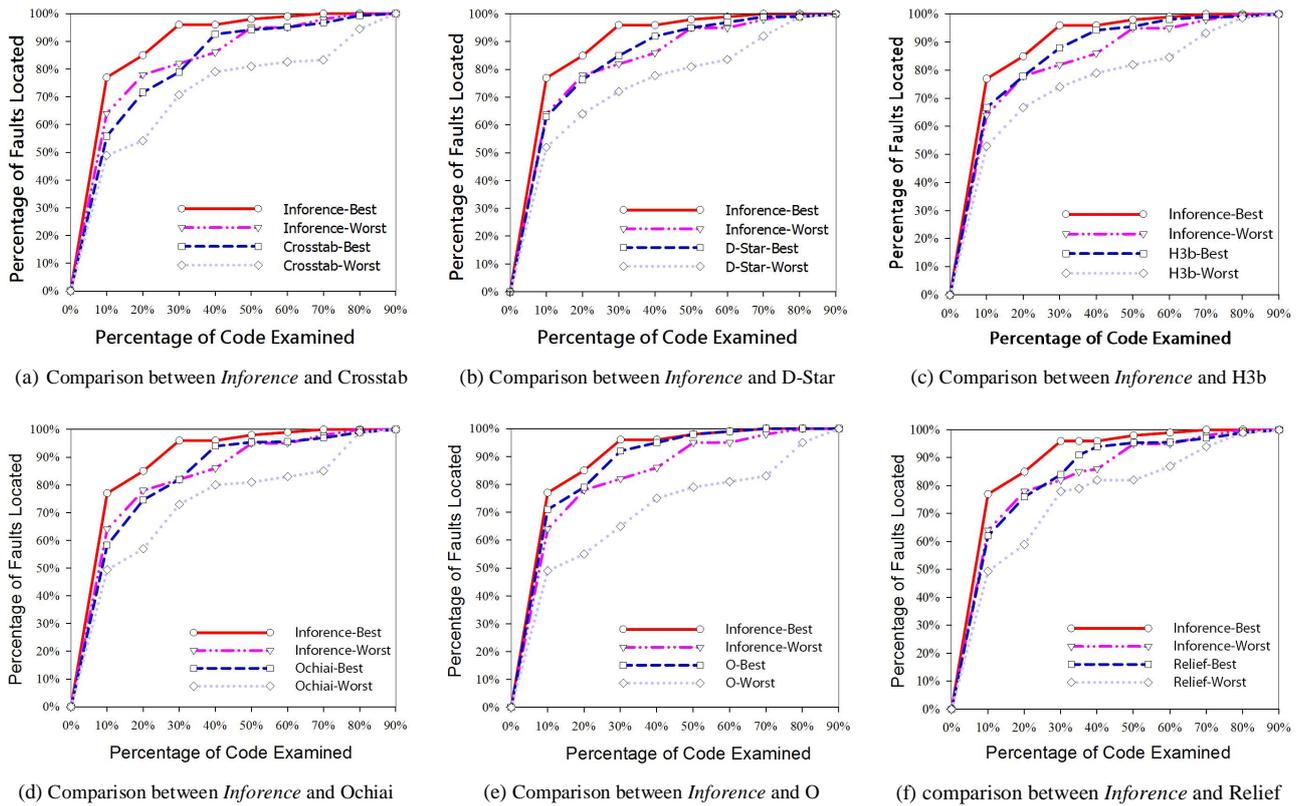

**Fig. 2.** *EXAM* score-based comparison on Siemens suite.

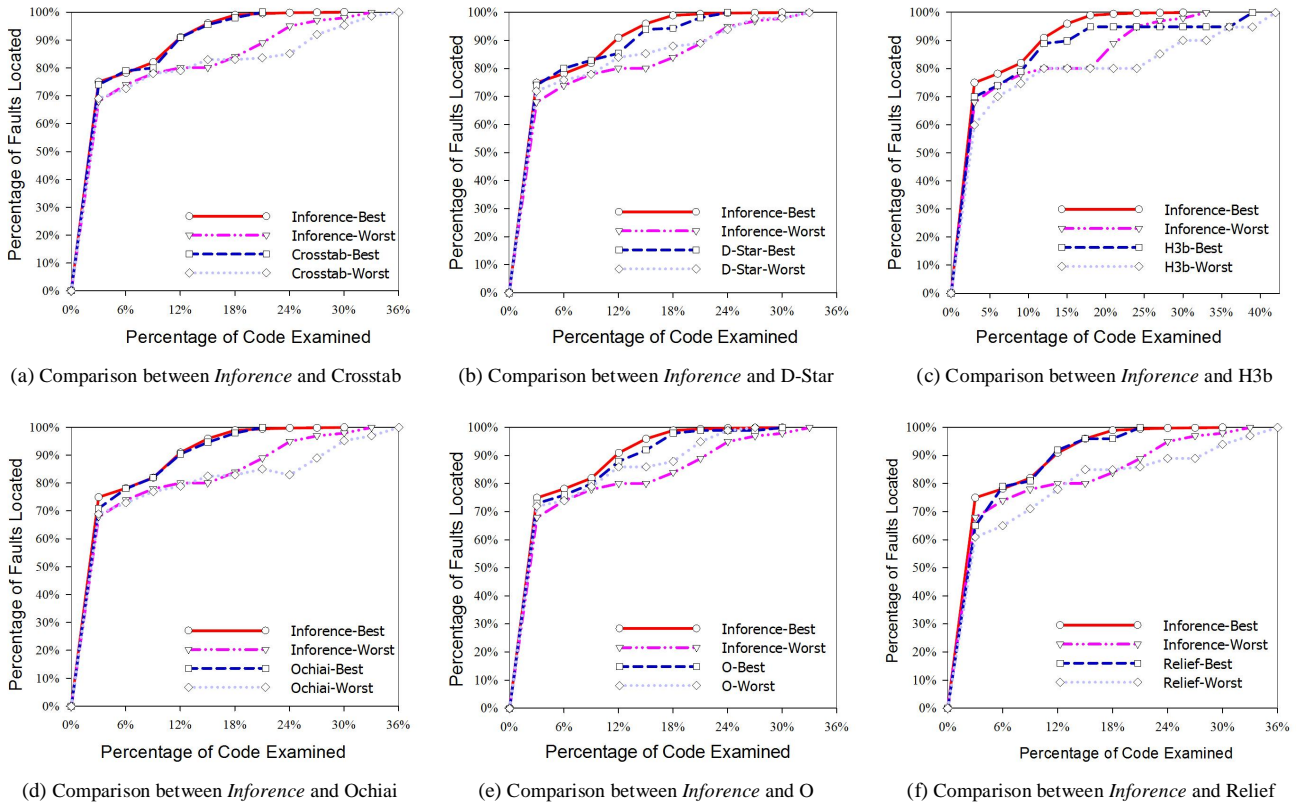

**Fig. 3.** *EXAM* score-based comparison on Sed program

The comparison between the effectiveness of *Inforence* and peer techniques on *Sed* program is shown in Fig. 3. We observe that the best effectiveness of *Inforence* is, in general, better than the best effectiveness of peer techniques, but its worst effectiveness is worse than the worst effectiveness of D-Star and O.

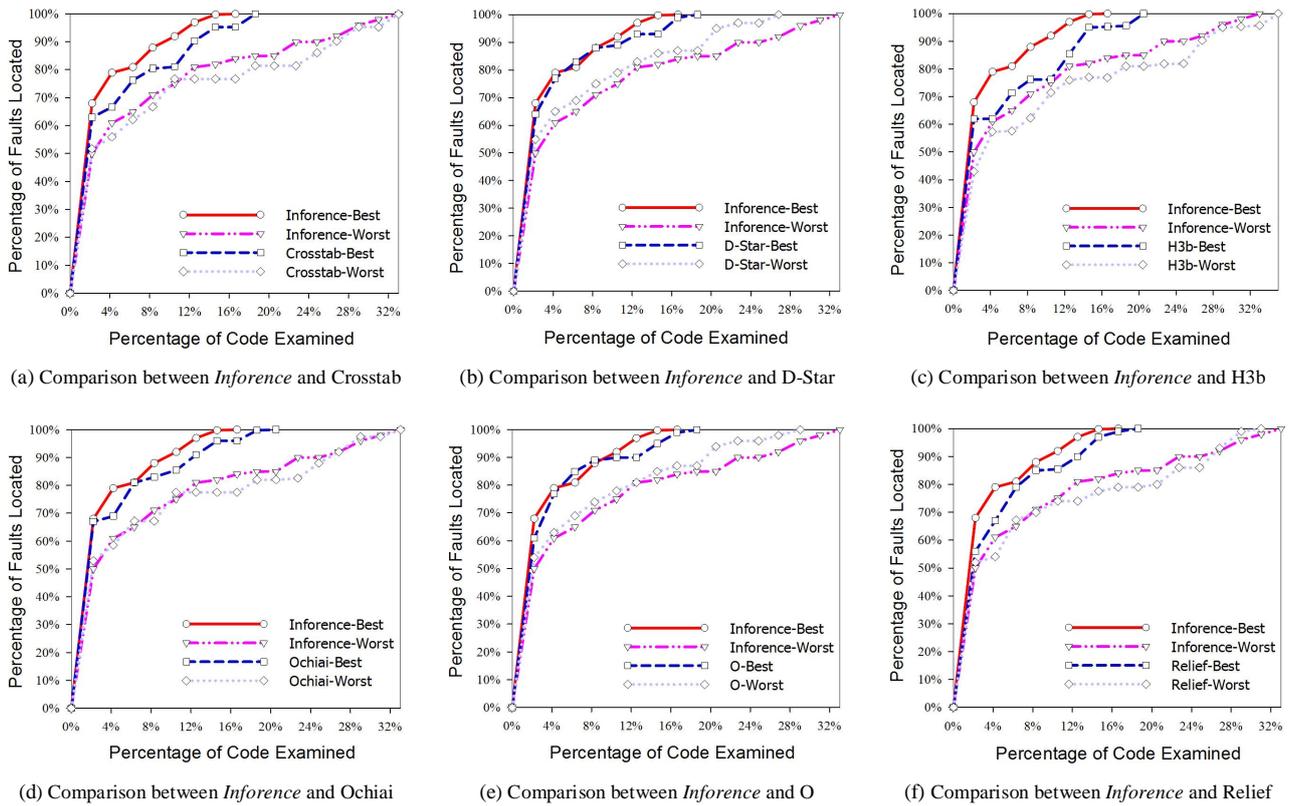

**Fig. 4**. *EXAM* score-based comparison on Gzip program

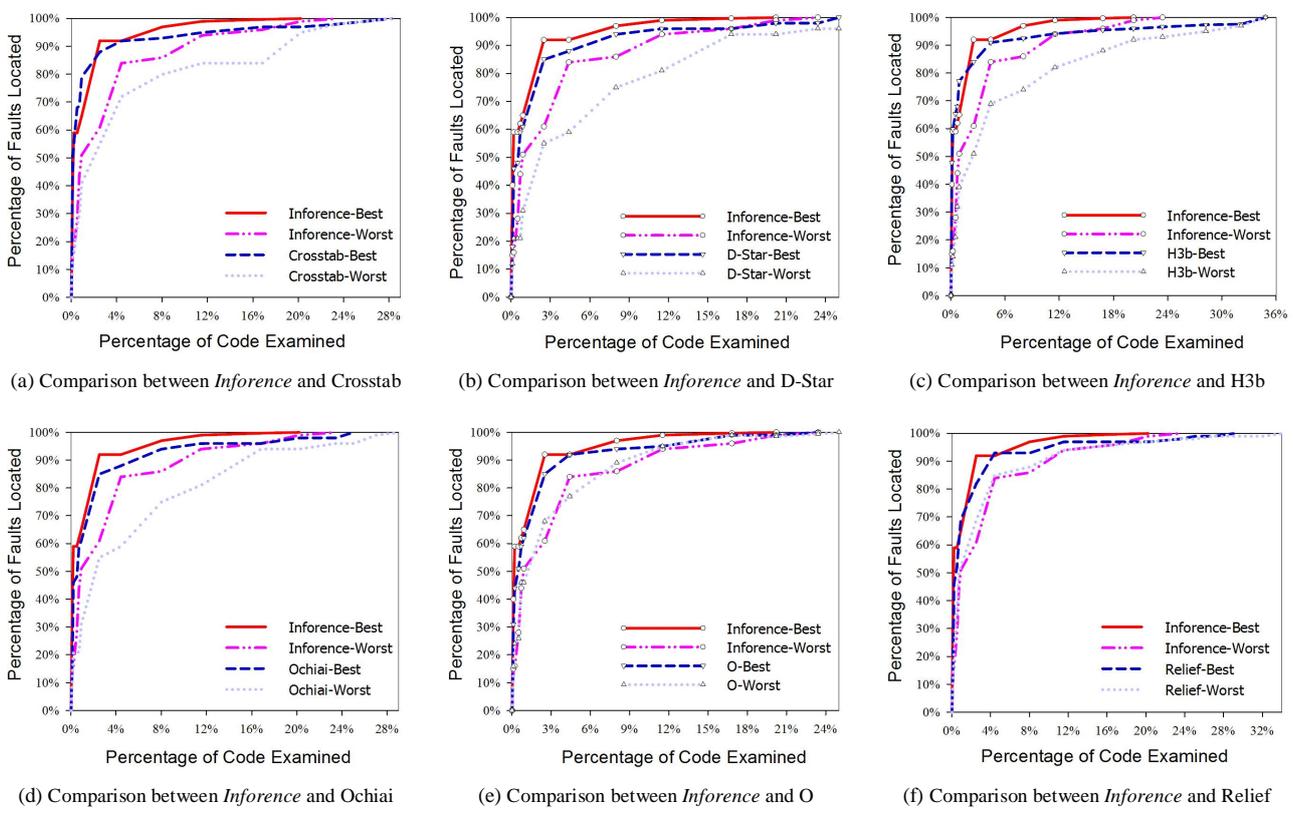

**Fig. 5**. *EXAM* score-based comparison on Space program

The comparison between the effectiveness of *Inforence* and peer approaches on *gzip* program is shown in Fig. 4. We observe that the effectiveness of *Inforence* is in general better than the effectiveness of peer techniques in both best and worst cases. The comparison between the effectiveness of *Inforence* and peer approaches on *space* program is shown in Fig. 5. We observe that the best effectiveness of *Inforence* is, in general, better than the best effectiveness of peer techniques, but its worst effectiveness is worse than the worst effectiveness of O.

4.3.2 Results on programs with multiple bugs

Thus far, the evaluation of *Inforence* has been with respect to programs that have exactly one fault. In this section, we discuss and demonstrate how *Inforence* may be applied to programs with multiple faults as well.

For programs with multiple faults, the authors of [47] define an evaluation metric, *Expense*, corresponding to the percentage of code that must be examined to locate the first fault as they argue that this is the fault that programmers will begin to fix. We note that the Expense score, though defined in a multi-fault setting, is the same as the *EXAM* score used in this paper. The Expense score is really part of a bigger process that involves locating and fixing all faults (that result in at least one test case failure) that reside in the subject program. After the first fault has successfully been located and fixed, the next step is to re-run test cases to detect subsequent failures, whereupon the next fault is located and fixed. The process continues until failures are no longer observed, and we conclude (but are not guaranteed) that there are no more faults present in the program. This process is referred to as the *one-fault-at-a-time* approach, and thus the Expense score only assesses the fault localization effectiveness with respect to the first iteration of the process.

A multiple fault version is a faulty version X with $k$ faults that is made by combining $k$ faulty versions from a set $\{x_1; x_2; \ldots; x_k\}$ where each bug i in X corresponds to the faulty version $x_i$. In practice, developers are aware of the number of failed test cases for their programs but are unaware of whether a single fault or many faults caused those failures. Thus, developers usually target one fault at a time in their debugging. Since there is more than one way to create a multi-bug version, using only one may lead to a biased conclusion. To overcome this problem, 30 distinct faulty versions with 3, 4, 5 and 6 bugs, respectively, for gzip, grep, sed, make, and space were created. Altogether, there are 600 multi-bug programs in our study. We apply *Inforence* and other representative techniques to faulty versions with multiple bugs using the *one-bug-at-a-time* method. Tables 8 and 9 give the average number of statements that need to be examined to find the first bug for the best and the worst cases. For example, the average number of statements examined by *Inforence* for the five-bug version of gzip is 36.87 for the best case and 61.59 for the worst, whereas only 36.45 (best) and 64.44 (worst) statements need to be examined by crosstab and 119.62 (best) and 182.5 (worst) by D-Star.

**Table 8** Average number of statements examined to locate the first bug (best case)

|  | Gzip | | | | Grep | | | | Sed | | | |
| --- | --- | --- | --- | --- | --- | --- | --- | --- | --- | --- | --- | --- |
|  | 3-bug | 4-bug | 5-bug | 6-bug | 3-bug | 4-bug | 5-bug | 6-bug | 3-bug | 4-bug | 5-bug | 6-bug |
| *Inforence* | 34.51 | 37.26 | 36.87 | **39.66** | 144.25 | **121.32** | **175.16** | 139.26 | 41.51 | 46.88 | 37.69 | 34.56 |
| D-Star | 101.68 | 98.51 | 119.62 | 126.41 | **135.66** | 219.56 | 281.82 | 112.43 | 62.51 | 69.85 | 172.66 | 195.41 |
| H3b | 71.52 | 69.21 | 88.6 | 62.7 | 211.65 | 277.48 | 289.55 | **80.26** | 188.21 | 246.59 | **23.30** | 29.52 |
| H3c | 79.26 | 78.85 | 88.1 | 69.42 | 231.46 | 285.04 | 271.00 | 79.63 | 181.51 | 209.17 | 25.26 | 42.18 |
| Crosstab | **25.66** | **26.78** | **36.45** | 40.12 | 246.25 | 291.51 | 355.25 | 196.63 | **35.85** | **28.62** | 36.25 | **22.62** |
| Ochiai | 105.88 | 112.56 | 138.75 | 151.88 | 189.76 | 258.56 | 281.73 | 126.11 | 51.65 | 59.45 | 276.96 | 355.6 |
| O | 55.12 | 53.89 | 53.58 | 51.65 | 289.65 | 276.5 | 272.11 | 176.98 | 58.45 | 61.82 | 61.26 | 58.41 |
| O$^P$ | 169.54 | 158.8 | 151.45 | 151.23 | 289.65 | 276.5 | 272.11 | 176.98 | 218.51 | 319.79 | 376.52 | 251.63 |
| Relief | 56.44 | 67.51 | 82.25 | 106.76 | 156.82 | 217.66 | 188.92 | 151.55 | 48.20 | 162.74 | 95.42 | 170.66 |
| Gen-Entropy | 39.14 | 49.75 | 38.11 | 52.45 | 162.82 | 129.85 | 192.64 | 148.46 | 42.15 | 56.70 | 41.65 | 36.10 |
| GP02 | 128.52 | 135.76 | 154.21 | 144.65 | 196.33 | 275.6 | 296.25 | 141.3 | 55.74 | 79.63 | 144.52 | 148.68 |
| GP03 | 95.41 | 106.5 | 114.64 | 102.5 | 174.24 | 248.62 | 214.83 | 147.66 | 42.85 | 67.3 | 92.78 | 82.9 |
| GP19 | 108.2 | 126.45 | 139.56 | 131.75 | 194.36 | 269.5 | 262.88 | 144.62 | 48.87 | 77.56 | 125.3 | 116.78 |
| Baah, 2011 | 79.20 | 126.55 | 141.65 | 149.31 | 194.28 | 270.56 | 236.50 | 117.76 | 74.56 | 62.55 | 166.82 | 146.24 |

**Table 8** continued

|  | Make | Space |
| --- | --- | --- |

|  | 3-bug | 4-bug | 5-bug | 6-bug | 3-bug | 4-bug | 5-bug | 6-bug |
|---|---|---|---|---|---|---|---|---|
| *Inforence* | **124.89** | **116.56** | 157.33 | **51.33** | **68.23** | **64.95** | **77.41** | **58.32** |
| D-Star | 201.52 | 185.36 | **151.21** | 64.25 | 121.85 | 142.85 | 110.52 | 101.81 |
| H3b | 381.8 | 360.50 | 358.45 | 106.12 | 192.52 | 178.58 | 198.45 | 166.11 |
| H3c | 460.12 | 491.56 | 521.6 | 146.11 | 216.46 | 236.40 | 271.52 | 156.88 |
| Crosstab | 181.55 | 165.26 | 136.9 | 51.63 | 109.45 | 101.58 | 89.51 | 75.65 |
| Ochiai | 211.28 | 199.53 | 156.89 | 65.21 | 136.59 | 119.26 | 105.17 | 88.63 |
| O | 206.50 | 250.59 | 278.11 | 108.46 | 195.85 | 209.82 | 221.74 | 159.66 |
| O$^P$ | 369.21 | 482.74 | 499.25 | 151.63 | 221.62 | 251.45 | 298.25 | 172.45 |
| Relief | 232.50 | 165.44 | 161.62 | 62.37 | 118.24 | 104.45 | 92.85 | 69.55 |
| Gen-Entropy | 148.45 | 136.20 | 174.38 | 62.90 | 77.52 | 81.25 | 85.33 | 70.22 |
| GP02 | 256.85 | 274.64 | 194.5 | 96.25 | 194.35 | 174.9 | 159.74 | 132.16 |
| GP03 | 218.56 | 208.41 | 169.24 | 78.6 | 152.15 | 138.64 | 122.7 | 94.28 |
| GP19 | 244.65 | 261.78 | 185.62 | 92.35 | 182.46 | 146.14 | 142.36 | 117.5 |
| Baah, 2011 | 236.7 | 184.35 | 181.85 | 102.45 | 115.26 | 107.66 | 96.44 | 92.82 |

**Table 9** Average number of statements examined to locate the first bug (worst case)

|  | Gzip | | | | Grep | | | | Sed | | | |
|---|---|---|---|---|---|---|---|---|---|---|---|---|
|  | 3-bug | 4-bug | 5-bug | 6-bug | 3-bug | 4-bug | 5-bug | 6-bug | 3-bug | 4-bug | 5-bug | 6-bug |
| *Inforence* | **58.46** | 74.48 | **61.59** | 104.18 | 241.21 | **229.45** | 308.45 | **114.94** | 62.69 | 54.82 | 58.10 | 44.94 |
| D-Star | 120.49 | 98.25 | 182.5 | 188.45 | 255.67 | 281.32 | **298.76** | 150.33 | 144.62 | 111.62 | 84.60 | 85.16 |
| H3b | 109.26 | 120.88 | 135.51 | 115.96 | 371.26 | 378.38 | 391.62 | 142.88 | 288.21 | 428.12 | 59.30 | 71.52 |
| H3c | 129.26 | 141.85 | 158.1 | 119.42 | 399.19 | 391.36 | 380.56 | 146.12 | 224.51 | 361.17 | 65.26 | 82.18 |
| Crosstab | 59.21 | **58.64** | 64.44 | **82.36** | 529.45 | 699.05 | 832.65 | 262.64 | **55.85** | **48.62** | **52.66** | **41.62** |
| Ochiai | 165.88 | 185.25 | 215.48 | 202.46 | **226.78** | 341.26 | 378.96 | 176.61 | 351.65 | 79.45 | 376.96 | 525.6 |
| O | 1529.1 | 1668.4 | 1679.58 | 1511.6 | 421.65 | 391.85 | 394.62 | 295.71 | 2056.5 | 2128.8 | 2346.8 | 3190.4 |
| O$^P$ | 289.54 | 295.8 | 261.48 | 242.38 | 421.65 | 391.85 | 394.62 | 295.71 | 618.51 | 519.79 | 576.52 | 701.63 |
| Relief | 141.69 | 226.43 | 115.25 | 139.46 | 259.78 | 291.45 | 322.58 | 139.75 | 194.66 | 84.50 | 169.21 | 325.68 |
| Gen-Entropy | 81.35 | 110.64 | 85.70 | 117.58 | 291.82 | 274.36 | 348.80 | 175.69 | 106.69 | 88.62 | 82.46 | 59.50 |
| GP02 | 195.74 | 244.5 | 263.24 | 251.4 | 294.1 | 366.6 | 451.64 | 250.8 | 399.62 | 112.32 | 482.7 | 645.45 |
| GP03 | 174.12 | 196.68 | 214.52 | 218.42 | 248.24 | 358.74 | 394.82 | 194.46 | 379.49 | 89.1 | 405.52 | 598.54 |
| GP19 | 185.25 | 224.45 | 241.2 | 255.35 | 272.85 | 369.52 | 425.2 | 240.51 | 385.42 | 106.24 | 453.44 | 628.27 |
| Baah, 2011 | 124.62 | 138.30 | 194.72 | 196.48 | 264.92 | 312.66 | 354.35 | 148.36 | 146.18 | 135.6 | 112.35 | 104.86 |

**Table 9** continued

|  | Make | | | | Space | | | |
|---|---|---|---|---|---|---|---|---|
|  | 3-bug | 4-bug | 5-bug | 6-bug | 3-bug | 4-bug | 5-bug | 6-bug |
| *Inforence* | **195.62** | **176.48** | 279.16 | 148.36 | **131.63** | **114.25** | **107.71** | **128.92** |
| D-Star | 271.50 | 179.36 | 297.98 | 160.33 | 191.65 | 226.25 | 189.42 | 194.21 |
| H3b | 461.8 | 490.50 | 478.45 | 198.12 | 282.52 | 242.58 | 278.45 | 256.51 |
| H3c | 660.12 | 621.56 | 669.6 | 265.11 | 296.46 | 286.40 | 341.82 | 250.18 |
| Crosstab | 421.55 | 445.26 | 326.9 | 401.63 | 189.45 | 201.58 | 179.51 | 145.65 |
| Ochiai | 291.28 | 309.53 | **252.89** | **132.21** | 206.59 | 210.26 | 167.14 | 148.63 |
| O | 5128.5 | 4911.6 | 4571.1 | 3651.3 | 315.85 | 329.82 | 297.25 | 258.66 |
| O$^P$ | 669.21 | 601.74 | 659.25 | 281.63 | 281.62 | 328.45 | 421.56 | 351.45 |
| Relief | 312.45 | 362.78 | 265.25 | 144.63 | 178.24 | 184.45 | 152.85 | 192.40 |
| Gen-Entropy | 228.50 | 196.44 | 311.52 | 164.26 | 146.82 | 186.60 | 164.23 | 142.85 |
| GP02 | 342.54 | 366.36 | 290.58 | 174.63 | 294.22 | 282.44 | 195.63 | 174.71 |
| GP03 | 310.9 | 341.47 | 284.72 | 154.45 | 252.18 | 236.51 | 165.74 | 152.4 |
| GP19 | 329.13 | 360.24 | 289.9 | 164.27 | 285.45 | 239.25 | 184.25 | 165.32 |
| Baah, 2011 | 348.54 | 296.22 | 318.46 | 186.36 | 168.74 | 202.65 | 199.42 | 186.32 |

We observe that, regardless of best or worst case, *Inforence* is very effective in the case of make and space programs, among all the competing techniques. In the case of other subject programs, *Inforence* is always one of the two most effective techniques. The presence of multiple bugs in a program may cause the fault localization method to not be able to distinguish faults predictors and it may hinder a fault's ability to be localized in the presence of other faults. In the case of some multiple bug settings, *Inforence* could not perform well in identifying bug-related statements and resultant cause-effect chains require greater amount of manual conde inspection.

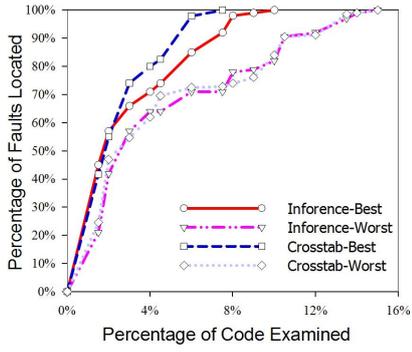
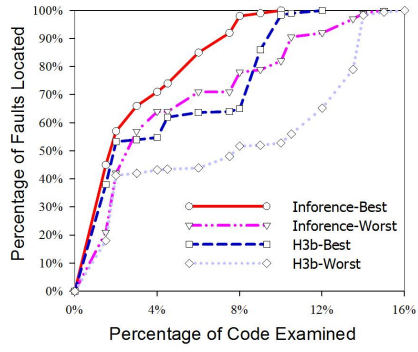
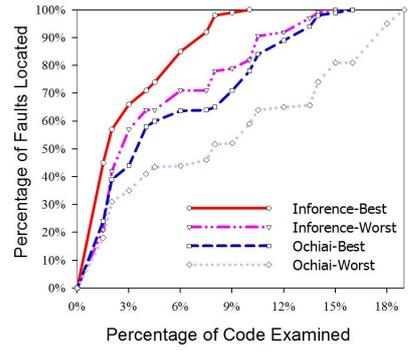

(a) Comparison between *Inforence* and Crosstab   (b) Comparison between *Inforence* and H3b   (c) Comparison between *Inforence* and Ochiai

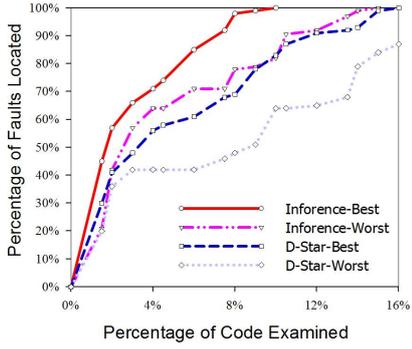
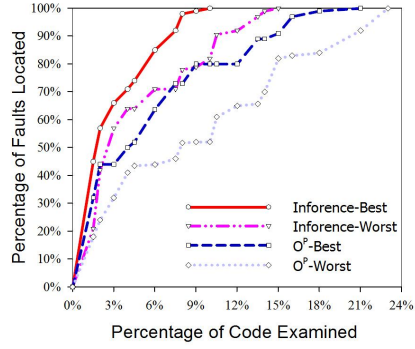
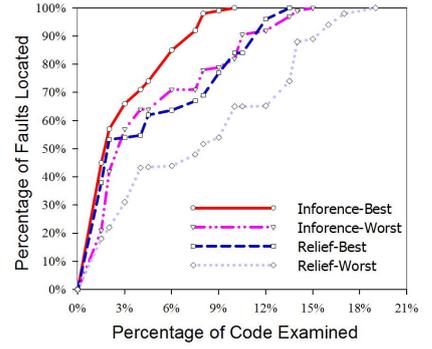

(d) Comparison between *Inforence* and D-Star   (e) Comparison between *Inforence* and $O^P$   (f) Comparison between *Inforence* and Relief

**Fig. 6**. EXAM score-based comparison on gzip program (Multiple-bug case)

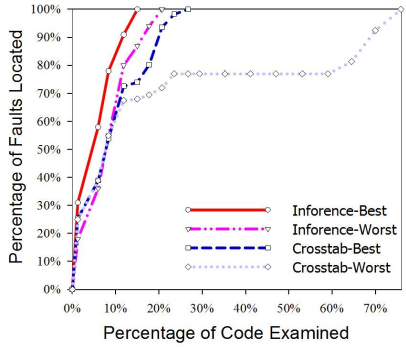
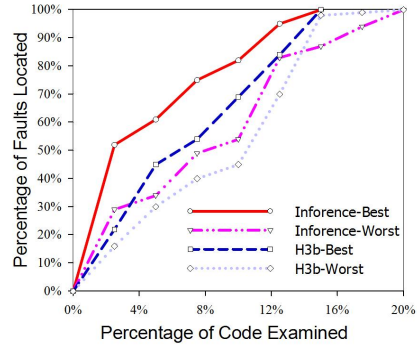
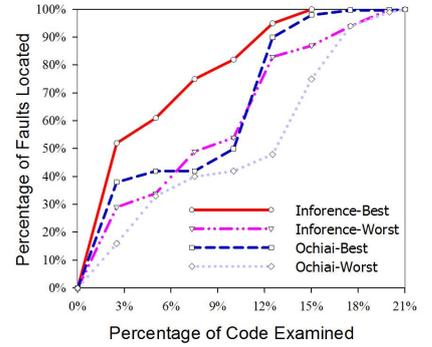

**(a)** Comparison between *Inforence* and Crosstab   **(b)** Comparison between *Inforence* and H3b   **(c)** Comparison between *Inforence* and Ochiai

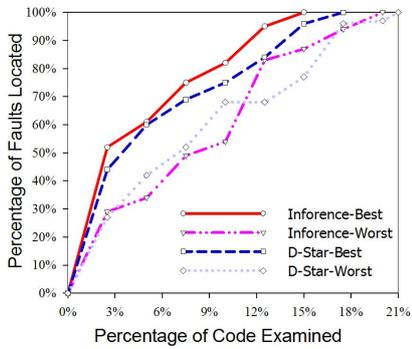
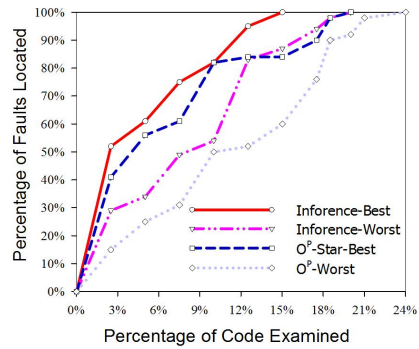
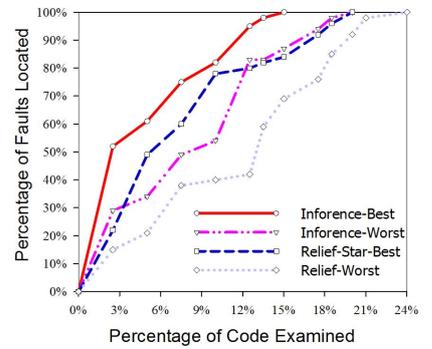

**(d)** Comparison between *Inforence* and D-Star   **(e)** Comparison between *Inforence* and $O^P$   **(f)** Comparison between *Inforence* and Relief

**Fig. 7**. *EXAM* score-based comparison on grep program (Multiple-bug case)

Fig. 6 and Fig. 7 present comparison between *Inforence* and representative techniques for gzip and grep,

respectively. From Part (a), for gzip, by examining 4 % of the code *Inforence* can locate the first bug of 82 (68.33 %) of 120 multi-bug versions in the best cases and 77 (64.16 %) in the worst; Crosstab can locate 96 (80 %) for best cases and 75 (62.5 %) for the worst. H3b can locate 67 (55.83 %) for best cases and 52 (43.33 %) for the worst; Ochiai can get 55 (45.83 %) for best cases and 46 (38.33 %) for the worst; D-Star can locate 65 (54.16 %) for best cases and 50 (41.66 %) for the worst.

Considering the grouping effect of statements on program failure helps programmers detect multiple bugs in programs. When a group of highly correlated statements, all together, affects the program failure, two scenarios can be considered:

1- Highly correlated statements have a combinatorial impact on program termination status and program failure may be revealed as the mutual effect of program statements on each other (in the case of programs with a single fault).

2- Highly correlated statements have a combinatorial impact on program failures caused by a single fault (in the case of programs with multiple faults). In this case, the faulty statements corresponding to all program faults can be chosen by proposed information theoretic analysis and the faulty statements corresponding to each individual fault are included into a cause-effect chain separated from chains including faulty statements of other existing faults of a program.

### 4.3.3 Providing failure context

Providing the fault propagation path for the purpose of discovering the location of faults is one of the expectations that programmers generally have from automatic fault localization methods. However, their expectations were not fulfilled by the most of existing fault localization methods. As mentioned earlier, a significant characteristic of *Inforence* is providing the context of failure in terms of cause-effect chains of program statements.

Since *Inforence* considers the grouping effect of program statements on program termination status during information theoretic analysis, high performance in providing cause-effect chains for program faults is expected. Cause-effect chains are provided to the programmer in the form of a sequence of statements. To evaluate the performance of *Inforence* in this case, precision, recall and F-measure metrics are used.

**Table 10** Performance of *Inforence* in providing failure context

| Program | Inforence | | |
|---------|-----------|--------|-----------|
| | Precision | Recall | F-Measure |
| Make | 84.8 | 91.2 | 87.9 |
| Sed | 82 | 88 | 84.9 |
| Bash | 86 | 94 | 89.8 |
| Space | 90 | 96 | 92.9 |
| Gzip | 82 | 92 | 86.7 |
| Grep | 92 | 95 | 93.5 |

Precision and recall are important performance measures. The higher the precision, the less effort wasted in testing and inspection; and the higher the recall, the fewer cause-effect chains go undetected. However, there is a trade-off between precision and recall [48]. For example, if the proposed method provides only one cause-effect chain for a fault that navigates the programmer to the root cause of the failure, the method's precision will be one. However, the method's recall will be low if there are other cause-effect chains. Therefore, F-measure is needed which combines recall and precision in a single efficiency measure [48]. Table 10 presents the results of conducted experiments on cause-effect chains with regards to precision, recall, and F-measure performance metrics.

### 4.3.4 Complexity analysis

Finally, we analyze the time complexity of the algorithm 1 and also present the time overhead of *Inforence* comparing with Ochiai, as a representative of SBFL approaches and Baah's causal approach [16] in Table11. The algorithm has linear complexity $O(n)$ to calculate the relevance $R(s, out)$ for each statement (Step 3), where $n$ is the number of statements in the *Potentially faulty candidate statements PS*. Assume that the number of remaining statements in $PS$ is $m$, which decreases from $n$ to $n - \delta$, then the time complexity of the $J(s)$ calculating procedure (Steps 5–7) and updating procedure (Steps 11–14) are $O(m)$ and $O(m-1)$ for each while loop. It has a worse-case complexity $O(n^2)$ when all statements are selected (it executes $\sum_{i=n}^{1} i + \sum_{j=n-1}^{1} j = n^2$ times in total). However, in general cases the threshold value δ is certainly much smaller, which makes a reasonable time complexity $O(n\delta)$ (i.e.,

$\sum_{i=n}^{n-\delta} i + \sum_{j=n-1}^{n-\delta} j$) for Algorithm1. Since the threshold value is set to $30\% \times |PS|$, the complexity of algorithm1 is $O(n \times |PS|)$.

The time overhead of *Inference* comparing with Ochiai and Baah's causal approach [16] is provided in Table 11. Column *preprocessing* illustrates the time overhead spent for all test cases by fault localization approaches, including the time required for computing slices for *Inference*. Column *Information theoretic analysis* presents the computational time of selecting the bug-related statements using the proposed algorithm. Column *suspiciousness computing* presents the computational time of suspiciousness, and the last two columns present the division of the total time required by *Inference*, Baah and Ochiai to illustrate the high time cost of our approach regarding to Ochiai and low time cost regarding to Baah's causal method. Column $\#Inference/\#Ochiai$ in Table 11 shows that the time overhead of *Inference* for all subjects are less than 1.89 times of Ochiai, as a representative of SBFL approaches, in our experiment.

**Table 11** Time cost of SBFL, Baah's causal approach and *Inference* spent on all subjects

| Subject | Approach | Preprocessing | Information theoretic analysis | Suspiciousness computing | Total | $\frac{\#Inference}{\#Baah}$ | $\frac{\#Inference}{\#Ochiai}$ |
|---|---|---|---|---|---|---|---|
| Siemens | Inference | 47,625s | 1845s | 13,421s | 52,891s | | |
|  | Baah | 42,826s | 0 | 21,469s | 64,295s | 0.82 | 1.16 |
|  | Ochiai | 42,826s | 0 | 2,724s | 45,550s | | |
| Gzip | Inference | 4,793s | 1621s | 3,269s | 6,683s | | |
|  | Baah | 3,926s | 0 | 11,521s | 15,447s | 0.43 | 1.61 |
|  | Ochiai | 3,926s | 0 | 211s | 4,137s | | |
| Grep | Inference | 5,514s | 1683s | 3,566s | 8,763s | | |
|  | Baah | 4,394s | 0 | 13,514s | 17,908s | 0.49 | 1.89 |
|  | Ochiai | 4,394s | 0 | 262s | 4,656s | | |
| Sed | Inference | 5,614s | 1665s | 3,962s | 8,241s | | |
|  | Baah | 4,027s | 0 | 10,822s | 14,849s | 0.55 | 1.93 |
|  | Ochiai | 4,027s | 0 | 230s | 4,257s | | |
| Space | Inference | 26,531s | 1741s | 4,322s | 29,594s | | |
|  | Baah | 23,852s | 0 | 14,621s | 38,473s | 0.77 | 1.22 |
|  | Ochiai | 23,852s | 0 | 326s | 24,178s | | |
| Make | Inference | 10,246s | 1762s | 3,156s | 10,524s | | |
|  | Baah | 8,268s | 0 | 12,356s | 20,264s | 0.52 | 1.21 |
|  | Ochiai | 8,268s | 0 | 389s | 8,657s | | |
| Bash | Inference | 5,849s | 1894s | 2,520s | 7,263s | | |
|  | Baah | 4,681s | 0 | 8,962s | 13,643s | 0.53 | 1.50 |
|  | Ochiai | 4,681s | 0 | 156s | 4,837s | | |

## 5 Related Works

In this section, we provide an overview of related fault localization techniques. We mainly focus on machine learning-based, spectrum-based and slicing-based fault localization techniques that are related to our approach. Machine learning techniques are recently used with the aim of debugging. In this context, the problem can be assumed as an attempt to learn or deduce the location of a fault based on input data such as program coverage data and the execution result of each test case. In [8] Wong et al. proposed a fault localization technique based on a back-propagation (BP) neural network. Since BP neural networks suffer from issues such as paralysis and local minima, Wong et al. [6] proposed another approach based on RBF (radial basis function) networks, which are less vulnerable to these problems and have a faster learning rate. Ascari et al. [51] use neural networks and support vector machines (SVM) for fault localization in the context of object-oriented programming. Context-aware statistical debugging [11] combines the methods of feature selection, clustering, and static control flow graph analysis to identify the failure context.

The application of information theory in the context of software fault localization is recent. In the context of traditional SBFL, with the best of our knowledge, Lucia et al. are the first to use information gain as a part of their research to measure and compare a number of well-known similarity measures [50]. Their study just applies these measures following the same idea as in Tarantula [51], i.e., by building a similarity measure using the underlying dichotomy matrix. Information theoretic concepts have recently been applied [52] to prioritize tests to enhance fault localization effectiveness when it may not be feasible to run all tests, which is inspired by the *test prioritization* problem in regression testing [53-55]. Yoo et al. in [55], used Tarantula as their base measure for suspiciousness together with Shannon entropy to prioritize the tests to be run, so that Tarantula localization can likely converge faster. To our knowledge, the paper [56] is the first work that defines a specialized information theoretic solution based on feature selection using probabilistic divergence for more effective SBFL. Latent divergence introduces a new concept of feature selection, which complements previous techniques, such as neural networks, SVMs, and tree-based classification. Roychowdhury et al. in [9], proposed a method that uses standard feature selection algorithms like RELIEF and its variants to find the failure-relevant statements. They showed that the statements with maximum information diversity point to most suspicious lines of code. They also proposed a family of generalized entropies and showed that using mutual information based on generalized entropies allows more accurate fault localization that traditional techniques [10].

A program spectrum details the execution information of a program from certain perspectives and can be used to track program behavior. SBFL techniques use program spectrum to indicate entities more likely to be faulty. D-Star [2], evaluates the suspiciousness of statements by modifying the Kulczynski coefficient to D-Star (i.e., by adding an exponent to its numerator). Abreu et al. [7] applied the Ochiai coefficient in software fault localization for locating single bugs. Comparing with Tarantula [51], their experiments indicated that the Ochiai coefficient consistently outperforms Tarantula. A later study by Abreu et al. [57] evaluated the effectiveness of Tarantula and other proposed methods. Their experiments indicated that Ochiai performs the best for statement coverage independent of test cases. In [8], Naish et al. proposed two techniques, $O$ and $O^P$. The technique $O$ is designed for programs with a single bug, while $O^P$ is better applied to programs with multiple bugs. Data from their experiments suggest that $O$ and $O^P$ are more effective than Tarantula and Ochiai for single-bug programs. However, for better performance in multi-bug programs, they proposed another technique named $O^P$. A statistical fault localization technique based on crosstab is proposed in [3]. The impact of how each additional failed (or successful) test case can help locate program faults is investigated in [4]. The authors conclude that the contribution of the identified failed test cases are stepwise decreasing. This conclusion is also applicable for successful tests. The techniques are named H3b and H3c, which are more effective than those heuristics due to an additional stipulation such that the total contribution from all the failed tests that execute a statement s should be more than the total contribution from all the successful tests that execute s if s has been executed by at least one failed test case.

Instead of manually designing fault localization formulas, Yoo generates a number of formulas using Genetic Programming [58]. The effectiveness of these formulas are then theoretically studied by Xie et al. [46]. They find that GP13, GP02, GP03, and GP19 are the best ones for fault localization. They make three assumptions: i) a faulty program has exactly one fault; ii) for any given single-fault program, there is exactly one faulty statement; and iii) this faulty statement must be executed by all failed tests. They also assume that the underlying test suite must have 100% statement coverage. Unfortunately, many of these assumptions are over-simplified and do not hold for real-life programs [59].

To highlight that the assumption made by Xie et al. is not valid for many settings, the authors in [60], compared the performance of five theoretically best SBFL formulas with popular SBFL formulas (Tarantula and Ochiai) and showed that a relatively small reduction in test coverage can significantly affect the performance of the theoretically best SBFL formulas.

Comparisons among different SBFL techniques are frequently discussed in recent studies [5][11]. However,

there is no technique claiming that it can outperform all others under every scenario. In other words, an optimum SBFL technique does not exist, which is supported by Yoo et al.'s study [61].

Most of the earlier debugging effort relied on slicing-based techniques to help programmers reduce the search domain to quickly locate bugs. In [62], Xiaolin et al. proposed a method using a hybrid spectrum of full slices and execution slices to improve the effectiveness of fault localization. Their approach firstly computes full slices of failed test cases and execution slices of passed test cases respectively. Secondly it constructs the hybrid spectrum by intersecting full slices and execution slices. Finally, it computes the suspiciousness of each statement in the hybrid slice spectrum and generates a fault location report with descending suspiciousness of each statement.

To reduce the size of dynamic slices, Gupta et. al. in [63], integrated dynamic backward slicing with the idea of delta debugging [34]. In their later work [64], a bidirectional dynamic slice is computed for a specific identified decision-making statement known as a critical predicate. The critical predicate is a conditional branch statement which is likely to be responsible for the failure execution and if an execution instance of that predicate is switched from one outcome to another, the program produces a desirable output. The bidirectional dynamic slice contains statements in both forward and backward slices of a critical predicate. In a similar work in [65], a value-replacement based method is proposed. In this method, a set of values which have been used in an execution instance of a statement in a failing execution is replaced with some other values to analyze whether the program result changes from incorrect to correct output. If that happens, the statement is marked as a faulty statement for a single bug program. Techniques that rely on critical predicate switching [64] or value replacement [65] may have scalability problems for large programs with huge amount of data values. In predicate switching case, there might be many predicates in a program and a failure output may be the cause of more than a single critical predicate. Thereby, identifying critical predicates among a huge number of predicates does not seem to be trivial and straightforward. Another limitation of slicing techniques is their dependence on a single failing execution. Hence, they only identify a fault that is related to that failure and cannot detect other unknown faults of a program. They are also incapable of finding multiple bugs. Furthermore, due to the nature of slicing, some types of bugs (e.g. missing code) are hard to capture [68].

# 6 Threats to validity

Like any empirical study, there are some threats to the validity of our experiment. There are three main types of threats to validity that affect our studies: internal, external, and construct. Threats to external validity arise when the results of the experiment are unable to be generalized to other situations. We would like to emphasize that like any other fault localization methods, the effectiveness of *Inforence* varies for different programs, bugs, and test cases. While it is true that the evaluation of the methods presented here is based on empirical data and therefore we may not be able to generalize our results to all programs; it is for this reason that we observed the effectiveness of the methods across such a broad spectrum of programs. Each of the subject programs varies greatly from the other with respect to size, function, number of faulty versions studied, etc. This allows us to have greater confidence in the applicability of our fault localization method to different programs and the superiority of the results. However, the nature of faults in code, in general, is highly diverse and complex, and therefore our results may not be representative for all possible programs. Threats to internal validity concern factors that might affect dependent variables without the researcher's knowledge. The implementations of the algorithms we used in our studies could contain errors. The packages we used in our studies are open source and has been used by other researchers for experimentation, which provides confidence that the algorithms in the package are stable. The coverage measurement tools (such as whether the runtime trace can be correctly collected even if a program execution is crashed due to a segmentation fault) and environments (including compilers, operating systems, hardware platforms, etc.) are also expected to have an impact. An important threat to the construct validity of debugging approaches is the adequacy of quality metric that is chosen to measure their reported results. To mitigate this threat, we use evaluation metrics which are widely used and suitable effectiveness measures in the literature of fault localization. However, these metrics may not reflect the real-world cost of manually localizing the faults since, for example, for the results presented in this paper, we assume that if a programmer examines a faulty statement, he/she will identify the corresponding fault. At the same time, a programmer will not identify a non-faulty statement as faulty. If such perfect bug detection does not hold, then the number of statements that need to be examined to identify a bug may increase. However, such a concern also applies to other fault-localization techniques, and therefore, this is a common limitation. Furthermore, some categories of faults are easier to be located than the others. For instance, a missing code fault is relatively difficult to be identified in comparison with a fault which is located in an assignment statement.

## 7 Concluding remarks

In this paper, a new fault localization algorithm, so-called *Inference*, is proposed. Fault localization algorithms are mainly evaluated by measuring the amount of code to be manually examined around their reported fault suspicious statements before the actual failure origin is located. Fault suspicious statements can be pinpointed, relatively more accurately, by analyzing the combinatorial effect of statements on a program's termination status. To minimize the search space to pinpoint failure origins, the statements appearing in backward dynamic slices are initially selected for consideration.

We try to overcome the disadvantage that traditional feature selection algorithms often ignore some features which have strong discriminatory power as a group but are weak as individuals and demonstrate the need for this consideration in software fault localization context. In this way, those faults affected by a combination of program statements are revealed. We use a new concept named correlation ratio (CR) to judge the relationship between each candidate statements and newly selected statement. Candidate statements with a positive value of CR are deem to be interdependent with the newly selected statement and their weight should be raised. Besides, statements falling in other relationships can also be identified and their weight will be adjusted correspondingly. For example, causal relationships may bring about real redundancy of statements in most cases [12-14]. In this paper, interdependent group means that statements as a group have much stronger discriminatory power than the sum of each individual in isolation. In fact, it is difficult to discover the interdependent groups from program execution data. *Inference* adopts a re-weighting technique in order to let the statements which are interdependent with the newly selected one have higher priority.

*Inference* makes causal inference on a small subset of program statements that included in resultant cause-effect chain(s) on program failure, identified through proposed information theoretic analysis. Thus, we can claim that *Inference* is a scalable and cost-effective causal fault localization method. *Inference* also finds the failure context by identifying cause-effect chains. Our experimental results show that our proposed approach outperforms many distinguished methods in localizing faults, including standard feature selection based method

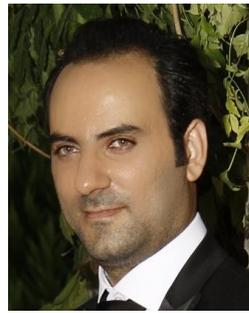

**Farid Feyzi** received the M.S. degree in Software Engineering from the Sharif University of Technology in 2012. He is currently a Ph.D. Candidate in the Department of Computer Engineering at Iran University of Science and Technology. His research focus is on developing statistical algorithms to improve software quality with an emphasis on statistical fault localization and automated test data generation.

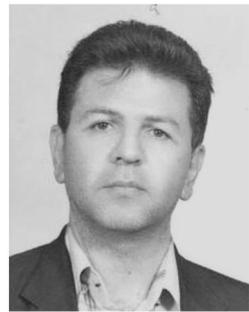

**Saeed Parsa** received his B.Sc. in Mathematics and Computer Science from Sharif University of Technology, Iran, his M.Sc. degree in Computer Science from the University of Salford in England, and his Ph.D. in Computer Science from the University of Salford, England. He is an associate professor of Computer Science at the Iran University of Science and Technology. His research interests include software engineering, software testing and debugging and algorithms.